\documentclass[acmsmall]{acmart}
\usepackage{multirow}
\usepackage{longfbox}
\usepackage{graphicx}
\usepackage{amsmath}
\usepackage[american]{babel}
\usepackage{microtype}
\usepackage{subcaption}
\usepackage[ruled,,linesnumbered ]{algorithm2e}
\usepackage{tcolorbox}
\usepackage{balance} 
\usepackage{tikz}
\usepackage{xcolor} 
\usepackage{enumitem} 
\usepackage{color}
\usepackage{url}
\definecolor{maroon}{RGB}{128,0,0} 
\definecolor{lightgreen}{RGB}{150,193,107}
\definecolor{lightorange}{RGB}{248,203,173}

\definecolor{confidence}{rgb}{0.0, 0.0, 0.0}
\definecolor{myblue}{RGB}{68, 84, 106}
\definecolor{markblue}{RGB}{63, 81, 181}
\definecolor{markred}{RGB}{197, 57, 41}

\newcommand*\rounded[1]{\tikz[baseline=(char.base)]{
        \node[shape=circle,inner sep=2pt,fill=lightgreen] (char) {\textcolor{white}{#1}}}}

\newcommand{\nltovis}{\textsc{NL2Vis \xspace}}%
\newcommand{\nltovisnospace}{\textsc{NL2Vis}}%

\AtBeginDocument{%
  }

\setcopyright{acmlicensed}
\acmJournal{PACMMOD}
\acmYear{2024} \acmVolume{2} \acmNumber{3 (SIGMOD)} \acmArticle{115} \acmMonth{6}\acmDOI{10.1145/3654992}

\begin{document}

\title{Automated Data Visualization from Natural Language via Large Language Models: An Exploratory Study}

\author{Yang Wu}
\authornote{Also with National Engineering Research Center for Big Data Technology and System, Services Computing Technology and System Lab, Cluster and Grid Computing Lab, School of Computer Science and Technology, Huazhong University of Science and Technology, Wuhan, 430074, China.}
\authornote{Both authors contributed equally to this research.}
\affiliation{%
 \institution{Huazhong University of Science and Technology}
 \country{China}
 }
\email{wuyang_emily@hust.edu.cn}

\author{Yao Wan}
\authornotemark[1]
\authornotemark[2]
\authornote{Yao Wan is the corresponding author.}
\affiliation{%
 \institution{Huazhong University of Science and Technology}
 \country{China}
 }
\email{wanyao@hust.edu.cn}

\author{Hongyu Zhang}
\affiliation{%
\institution{Chongqing University}
\country{China}
}
\email{hyzhang@cqu.edu.cn}

\author{Yulei Sui}
\affiliation{%
\institution{University of New South Wales} 
\country{Australia}
}
\email{y.sui@unsw.edu.au}

\author{Wucai Wei}
\authornotemark[1]
\affiliation{%
\institution{Huazhong University of Science and Technology}
 \country{China}
  }
  \email{m202273789@hust.edu.cn}

\author{Wei Zhao}
\authornotemark[1]
\affiliation{%
\institution{Huazhong University of Science and Technology}
 \country{China}
}
\email{m202073277@hust.edu.cn}

\author{Guandong Xu}
\affiliation{%
\institution{University of Technology Sydney} 
\country{Australia}
}
\email{guandong.xu@uts.edu.au}
 

\author{Hai Jin}
\authornotemark[1]
\affiliation{%
 \institution{Huazhong University of Science and Technology}
 \country{China}
 }
\email{hjin@hust.edu.cn}

\renewcommand{\shortauthors}{Yang Wu et al.}

\begin{abstract}
    The \textit{Natural Language to Visualization (\nltovisnospace)} task aims to transform natural-language descriptions into visual representations for a grounded table, enabling users to gain insights from vast amounts of data. 
    Recently, many deep learning-based approaches have been developed for \nltovisnospace.
    Despite the considerable efforts made by these approaches, challenges persist in visualizing data sourced from unseen databases or spanning multiple tables.
    Taking inspiration from the remarkable generation capabilities of \textit{Large Language Models (LLMs)}, this paper conducts an empirical study to evaluate their potential in generating visualizations, and explore the effectiveness of in-context learning prompts for enhancing this task.
    In particular, we first explore the ways of transforming structured tabular data into sequential text prompts, as to feed them into LLMs and analyze which table content contributes most to the \nltovisnospace. 
    Our findings suggest that transforming structured tabular data into programs is effective, and it is essential to consider the table schema when formulating prompts.
    Furthermore, we evaluate two types of LLMs: finetuned models (e.g., T5-Small) and inference-only models (e.g., GPT-3.5), against state-of-the-art methods, using the \nltovis benchmarks (i.e., nvBench). 
    The experimental results reveal that LLMs outperform baselines, with inference-only models consistently exhibiting performance improvements, at times even surpassing fine-tuned models when provided with certain few-shot demonstrations through in-context learning.
    Finally, we analyze when the LLMs fail in \nltovisnospace, and propose to iteratively update the results using strategies such as chain-of-thought, role-playing, and code-interpreter.
    The experimental results confirm the efficacy of iterative updates and hold great potential for future study. 
\end{abstract}

\ccsdesc[500]{Human-centered computing~Empirical studies in visualization}

\keywords{Data Visualization, Data Analysis, Natural Language Processing, Code Generation, Large Language Models, Exploratory Study}


\received{October 2023}
\received[accepted]{January 2024}

\maketitle

\section{Introduction}
Data visualizations, typically presented as charts, plots, and histograms, offer an effective means to represent, analyze, and explore data, as well as enable the identification and communication of valuable insights. 
Despite the availability of numerous tools (e.g., Tableau's Ask Data~\cite{setlur2019inferencing} and Amazon's QuickSight~\cite{quicksight}) and domain-specific programming languages (e.g., Vega-Lite~\cite{satyanarayan2016vega} and ggplot2~\cite{villanueva2019ggplot2}) for data visualization, crafting effective data visualizations remains a complicated effort for a range of users, particularly those with limited or no prior visualization experience.
Moreover, there is a pressing demand for visualizing data on smart devices such as tablets and mobile phones without requiring users to acquire data visualization expertise. 

To facilitate users in conducting data analytics, there has been an increasing interest in the automated generation of data visualizations from natural-language descriptions, denoted as \nltovisnospace~\cite{luo2021synthesizing,luo2021nvbench}.
Existing approaches to \nltovis mainly fall into two  categories: the rule-based~\cite{gao2015datatone,setlur2016eviza,hoque2017applying} and deep-learning-based~\cite{luo2018deepeye,luo2021synthesizing,luo2021nvbench,luo2021natural}.
DataTone~\cite{gao2015datatone}, Eviza~\cite{setlur2016eviza}, and Evizeon~\cite{hoque2017applying} employed a parser (i.e., the Stanford Core NLP Parser~\cite{manning2014stanford}), along with a set of predefined rules, to translate natural-language descriptions into visualization queries.
DeepEye~\cite{luo2018deepeye} introduced a novel approach that enables the generation of visualizations based on keyword queries, akin to search engine functionality.
Current methods for deep-learning-based techniques rely mostly on the encoder-decoder paradigm, which, in an end-to-end fashion, encodes the natural-language specification into hidden states and subsequently generates visualization queries.

\nltovis is similar to the the task of \textsc{NL2SQL} (also referred to as \textsc{Text2SQL})~\cite{kim2020natural}, wherein the objective is to translate natural-language descriptions into \textit{Structured Query Language (SQL)} queries.
Generally, the visualization is articulated using the \textit{Visualization Query Language (VQL)}, a language that shares similarities with SQL.
Both \nltovis and \textsc{NL2SQL} are based on input tables with diverse structures and aim to generate queries of various complexities, including selection, comparison, aggregation, and join operations.
In comparison to \textsc{NL2SQL}, one distinction is that \nltovis faces the additional challenge of considering intricate visualization attributes during generation, including the selection of chart types (e.g., bar, pie, line, and scatter).
Drawing inspiration from an established \textsc{NL2SQL} dataset (e.g., Spider~\cite{yu2018spider}), \citet{luo2021synthesizing} proposed the creation of a paired dataset for \nltovisnospace. 
Based on this dataset, a benchmark called nvBench~\cite{luo2021nvbench} is built and a Transformer-based model (named ncNet~\cite{luo2021natural}) is introduced.

\noindentparagraph{\textbf{\textup{LLMs for \nltovisnospace.}}}
Recently, \textit{Large Language Models (LLMs)}, such as GPT-3.5~\cite{openai2023gpt35} and LLaMA~\cite{touvron2023llama}, have demonstrated impressive capabilities for few-shot learning in many \textit{Natural Language Processing (NLP)} tasks, including question answering~\cite{tan2023evaluation,guo2023images}, machine translation~\cite{lyu2023new,wang2023document,ouyang2023llm}, and code generation~\cite{zhong2023study,liu2023your}.
As LLMs sequentially process the entire large-scale training dataset during the pre-training phase, they face a limitation in directly handling structured data, including tabular data.
Alternatively, we can serialize the tabular data, input it into the LLMs, and prompt the LLMs to generate data visualizations, which are in the form of domain-specific query language.
To fill this gap, this paper aims to address the following research question: ``\textit{Can LLMs be utilized for automating data visualization from natural-language descriptions grounded on a table and how?}''.

\begin{figure}[!t]
\centering
\includegraphics[width=0.99\textwidth]{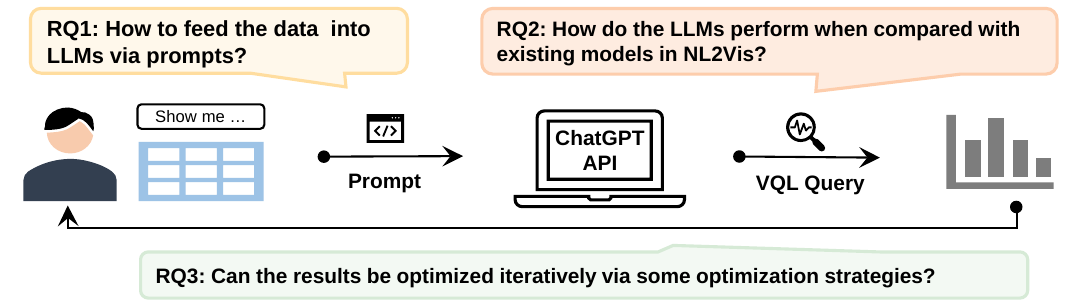}
\caption{An illustration of \nltovisnospace. 
The framework presents our investigation into prompt engineering (RQ1), overall performance (RQ2), and iterative updating (RQ3).
}
\label{fig_introduction}
\end{figure}

The challenges of leveraging the LLMs to automate data visualization from natural language are twofold.
\noindent\textbf{C1: Feeding the structural table into LLMs.} 
Given that LLMs exclusively accommodate sequential prompts, converting a structured grounded table into sequential prompts while preserving semantics poses a notable challenge.
Moreover, LLMs are recognized to face limitations due to their restricted token length. Consequently, they are unable to process entire extensive tables, making it challenging to comprehend comprehensive tabular information on a global scale.
\noindent\textbf{C2: Iteratively updating via conversation.} 
In contrast to traditional neural models for \nltovis that generate the data visualizations in a single attempt, one notable advantage of LLMs lies in their capacity to iteratively refine predicted outputs during conversations. Adapting traditional neural models to the new conversational paradigm also poses a significant challenge.

\noindentparagraph{\textbf{\textup{Our Work.}}}
To answer the aforementioned question, we conduct a pioneering empirical study to evaluate the capabilities of LLMs (i.e., T5~\cite{shaw2020compositional} and GPT-3.5~\cite{openai2023gpt35}) in automating data visualization from natural-language descriptions, comparing them with traditional approaches. 
Specifically, we structure the empirical study around the following three \textit{Research Questions (RQs)}, as depicted in Figure~\ref{fig_introduction}.  

\smallskip
\noindent\textbf{{RQ1: How to feed the natural-language query as well as the structural table into LLMs via prompting?}}
In this RQ, we first (1) explore the ways (i.e., table serialization, table summarization, table markup formatting, and table programming) to convert structured tabular data into sequential prompts, and subsequently (2) investigate which table content contributes most to the \nltovis in prompting.

\smallskip
\noindent\textbf{{RQ2: How do the LLMs perform when compared with several existing models in \nltovisnospace?}}
In this RQ, we first (1) evaluate the performance of LLMs (i.e., finetuned models such as T5-Small and T5-Base, and inference-only models such as \texttt{text-davinci-002} and \texttt{text-davinci-003}) for \nltovisnospace, against several traditional neural networks (i.e., \textsc{Seq2Vis}~\cite{luo2021synthesizing}, Transformer~\cite{vaswani2017attention}, ncNet~\cite{luo2021natural}, RGVisNet~\cite{song2022rgvisnet}), both under the in-domain and cross-domain settings, and (2) analyze how the number of in-context demonstrations affects the performance of LLMs.

\smallskip
\noindent\textbf{{RQ3: Can the results be iteratively updated via some optimization strategies?}}
In some cases, we observe that LLMs may fail to generate the correct visualization in a single attempt. In this RQ, we first (1) analyze when the LLMs fail in generating data visualization, and (2) propose to iteratively update the results via optimization strategies such as \textit{Chain-of-Thought (CoT)}, role-playing, self-repair, and code-interpreter.

\noindentparagraph{\textbf{\textup{Key Findings and Implications.}}}
In this paper, we observe the following important findings:
(1) To feed the structural table into LLMs, converting it into programs is an effective way. This finding inspires us to design programming patterns to encode tables. Additionally,  the schema information is sufficient for \nltovis task, which facilitates exploration of how to encode extra large databases with limited input length to LLMs.
(2) LLMs demonstrate significantly superior performance compared to traditional neural models for \nltovisnospace, highlighting their great potential under both in-domain and cross-domain settings. With in-context learning of LLMs, the demonstrations drawn from diverse tables can further increase performance. 
(3) The failure results can be further optimized via several iterative optimization strategies, such as CoT, role-playing, self-repair, and code-interpreter. 
To advance the visualization capability of LLMs, crafting multi-turn dialog prompts for automated optimization offers a promising prospect.


\noindentparagraph{\textbf{\textup{Contributions.}}}
The key contributions of this paper are as follows.
\setlist[itemize]{left=0pt}
\begin{itemize}
    \item  To the best of our knowledge, 
    this paper reports the first empirical study to investigate the capability of LLMs in automating data visualization from natural-language descriptions. Additionally, a benchmark of LLMs for \nltovis is built for further study.
    \item This paper systematically studies how to feed the data to visualize into the LLMs via prompts, and
    explores several optimization strategies for iteratively improving the failure results.
    \item To facilitate further study for other researchers, we release all the experimental data and source code used in this paper at \texttt{\url{https://github.com/CGCL-codes/naturalcc/tree/main/examples/explore-LLMs-for-NL2Vis}}~\cite{wan2022naturalcc}.
\end{itemize}

\noindentparagraph{\textbf{\textup{Organization.}}} 
The rest of this paper is structured as follows. 
We first introduce some background knowledge that will be used in this paper in Sec.~\ref{sec_background}. 
We then introduce our pipeline for \nltovis task in Sec.~\ref{llms_nl2vis} and the evaluation setup in Sec.~\ref{evaluation_setup}. 
Sec.~\ref{results} reports the experiment results with comprehensive analysis.
Sec.~\ref{discussion} is dedicated to discussing the potential threats to validity and the broader impacts of this paper.
We review the related work to this paper in Sec.~\ref{related_work}, and conclude this paper in Sec.~\ref{conclusion}.

\section{Background}
\label{sec_background}
In this section, we present foundational concepts of visualization query languages and LLMs, essential for understanding our work.

\subsection{Visualization Query Language (VQL)}
\label{vql}
In the realm of data visualization, one widely used grammar is Vega-Lite~\cite{satyanarayan2016vega}, which offers a concise and declarative JSON syntax for creating a diverse range of expressive visualizations suitable for data analysis and presentation. 
While Vega-Lite is intuitive and straightforward to use, training a sequence-to-sequence model to automatically generate hierarchical outputs, such as JSON format for Vega-Lite specifications, is challenging. Conversely, training a sequence-to-sequence model to generate sequential outputs is relatively more manageable.
In response to this challenge, several works~\cite{luo2018deepeye,luo2021synthesizing} introduce the 
VQL, which empowers users to articulate their data visualization requirements in a structured and efficient manner.
In contrast to Vega-lite, VQL queries remove 
structure-aware symbols such as parentheses, commas, and quotes, effectively transforming a JSON object into a sequence of keywords. This streamlining greatly simplifies the process of generating VQL queries.
Moreover, having eliminated all language-specific configurations, VQL is deemed language-agnostic that encompasses necessary data components (e.g., data operations) and vital visualization formats (e.g., visualization types). The VQL query could be easily transformed to diverse specifications (e.g., Vega-Lite, and ggplot2). 

\begin{table}[t!]
    \centering
    \caption{The visualization query language~\cite{luo2018deepeye}.}
    
    \begin{tabular}{ll}
        \hline
        \texttt{VISUALIZE}           &   $TYPE(\epsilon\{bar,pie,line,scatter\})$           \\
        \texttt{SELECT}            &        $X^{'}, Y^{'} (X^{'} \epsilon \{X, \texttt{BIN}(X)\}, Y^{'} \epsilon \{Y , \texttt{AGG}(Y)\})$      \\
        \texttt{FROM}   &   $D_1$           \\
        \texttt{JOIN}             & \( D_1 \bowtie D_2 \) \\
        \texttt{WHERE}   &        $X'$ $OP$ $v$      \\
        \texttt{TRANSFORM} & $X \rightarrow f(X) \, \text{where} \, f \in \{\texttt{BIN}, \texttt{GROUP}\}$\\
        \texttt{ORDER BY}              &  $X',Y'$          \\
        \texttt{AND/OR}           & \( \phi_1 \land \phi_2 \ ,\  \phi_1\lor  \phi_2 \) \\
        \texttt{Nested}           & \( Q(\text{subquery}) \) \\
        \hline
    \end{tabular}
    \label{tab_vql}
\end{table}

Table~\ref{tab_vql} shows the details of VQL~\cite{luo2018deepeye} for specifying visualization queries. 
Specifically, ``\texttt{VISUALIZE}'' specifies the visualization type, including bar, line, scatter, and pie. 
``\texttt{SELECT}'' specifies the chosen columns, 
where $X'$ represents either the original $X$ or its corresponding binned values,
and $Y'$ denotes either the original variable $Y$ or its aggregated value, derived by operations of \texttt{SUM}, \texttt{AVG}, and \texttt{COUNT}.
``\texttt{FROM}'' specifies the originating table.
``\texttt{JOIN}'' operation links tabular data from more than one table.
``\texttt{WHERE}'' filters the values that meet a certain condition, for instance, those greater than 10.
``\texttt{TRANSFORM}'' modifies the chosen columns, typically by binning $X$ into designated buckets or applying a \texttt{GROUP} operation.
``\texttt{BIN}'' operation divides the temporal data for visualization into several intervals, e.g., binning by year.  
``\texttt{GROUP}'' operation transforms the data into specific groups based on detailed stacked type or classification color.
``\texttt{ORDER BY}'' arranges the selected column in a particular sequence.
``\texttt{AND/OR}'' operation filters data based on multiple conditions.
``\texttt{Nested}'' operations nesting subquery, implements more complex data queries for visual representation.

\smallskip
\noindent\textbf{\textsc{Example 1.}}
Considering a natural-language query: \textit{List the name of technicians whose team is not ``NYY'', and count them by a bar chart, rank x-axis in ascending order}, the corresponding VQL query is as follows.
\begin{tcolorbox}[colback=white, colframe=black, sharp corners, left=1pt,right=1pt, top=1pt, bottom=1pt,boxrule=1pt]
\textbf{\texttt{VISUALIZE}} \textcolor{myblue}{\textit{bar}} \textbf{\texttt{SELECT}} \textcolor{myblue}{\textit{name}} , \textcolor{myblue}{\textit{COUNT(name)}} \textbf{\texttt{FROM}} \textcolor{myblue}{\textit{technician}} \textbf{\texttt{WHERE}} \textcolor{myblue}{\textit{team != ``NYY''}} \textbf{\texttt{GROUP BY}} \textcolor{myblue}{\textit{name}} \textbf{\texttt{ORDER BY}} \textcolor{myblue}{\textit{name asc}}
\end{tcolorbox}
We can see that this example is to select a name with its count from the technician table, where team $!=$ ``NYY''. Then, it groups the results by name with an ascend order, and finally visualizes the results using a bar chart.

\subsection{Large Language Models}
Over the past year, we have observed a growing proliferation of LLMs, including GPT-3~\cite{NEURIPS2020_1457c0d6} and LLaMA~\cite{touvron2023llama}. These LLMs have spearheaded a revolution in the field of NLP, especially in the related tasks of text generation~\cite{lyu2023new,wang2023document} and code generation~\cite{ouyang2023llm,zhong2023study,liu2023your}.
As pre-training LLMs on a large-scale dataset is a time-consuming and computationally expensive process (e.g., ChatGPT requires approximately 4,000 GPU hours for pre-training, at an estimated cost of 2 million dollars~\cite{gpt-cost}), numerous prompting techniques (e.g., in-context learning and chain-of-thought) have been developed with the goal of maximizing the effective utilization of LLMs.
\begin{figure}[!t]
\centering
\includegraphics[width=0.75\textwidth]{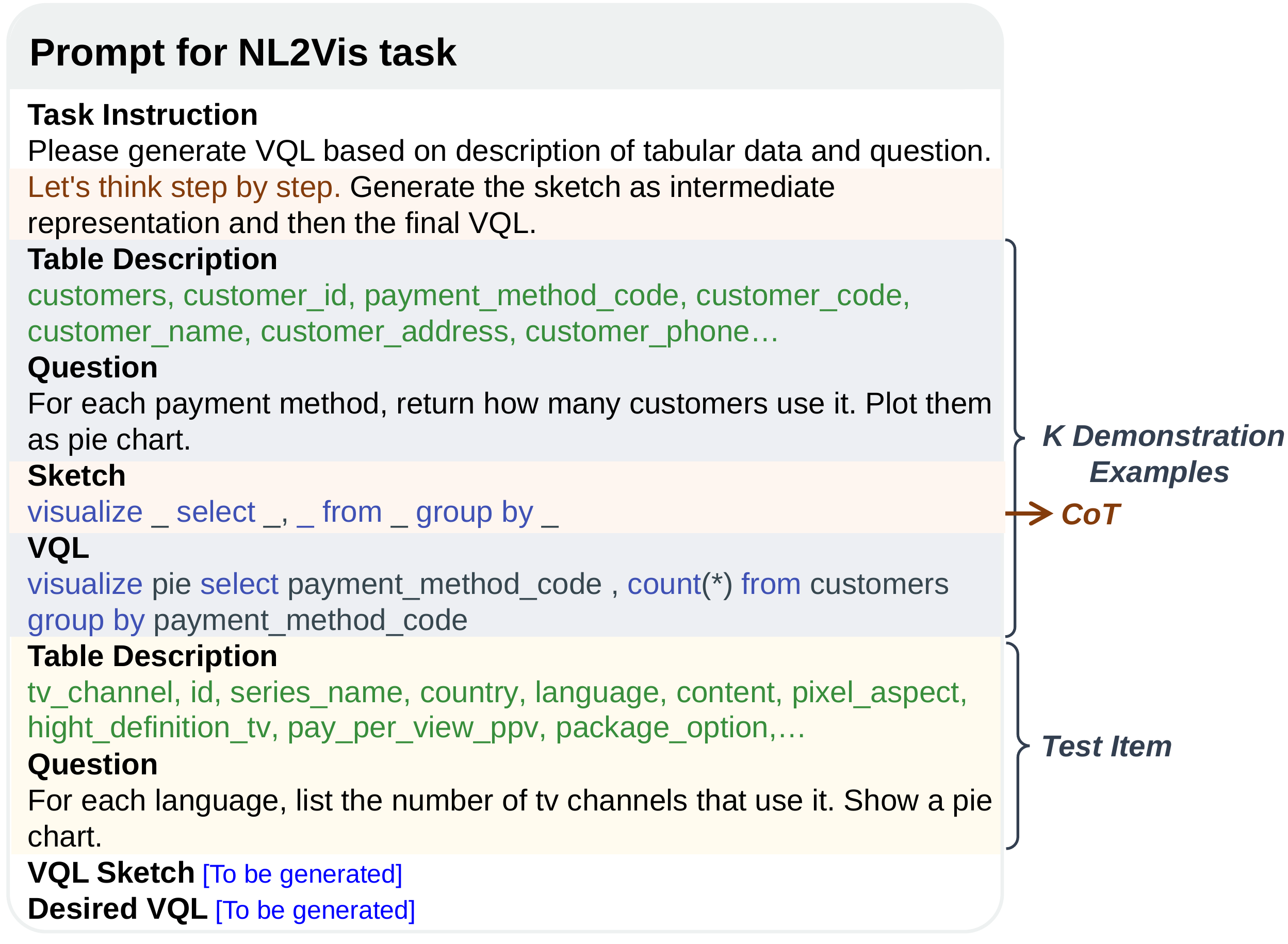}
\caption{An example to illustrate the usage of prompt in in-context learning of LLMs, the $k$ demonstration examples, the test item, and the CoT prompt.
}
\label{fig_prompt_construction}
\end{figure}

\subsubsection{Prompting}
With the emergence of LLMs, the learning paradigm is undergoing a transformation from the conventional ``\textit{pre-train and fine-tune}'' paradigm to a more innovative ``\textit{pre-train, prompt, and predict}'' framework~\cite{liu2023pre}.
In this new paradigm, 
instead of extensively fine-tuning LLMs to accommodate various downstream tasks, there is a shift towards reformulating these tasks to align more closely with the tasks for which LLMs are initially trained, with textual prompts guiding the process.
For instance, when assessing the emotion in a 
customer review like ``\textit{I like the book I have read today.}'', we may proceed with a prompt like ``\textit{It was\underline{\ \ \ \ }.}'' and 
request that LLM complete the sentence with an emotion-laden word.
In our specific scenario, involving a table and a natural-language query, we can construct a prompt as follows: ``\texttt{[TABLE]}, \texttt{[QUESTION]}, \textit{please generate VQL based on the description of tabular data and question.}'', where \texttt{[TABLE]} signifies the structured tabular data, and \texttt{[QUESTION]} represents the natural-language query provided by the end-user for exploring the data.

\subsubsection{In-Context Learning (ICL)}
ICL is a special form of prompt-based learning that leverages demonstration examples in prompts to promote the model's performance.
In ICL, in addition to describing the current question in the prompt, a few demonstration examples that have the same form as the question are also included in the prompt.
For instance, to generate a VQL query to count the number of television channels based on the language of each individual channel, a demonstrative example of counting customers by their preferred payment method is provided.
The model is expected to learn the pattern hidden in the demonstration, infer downstream tasks from examples, and accordingly make the right prediction.

Specifically, given a task instruction $I$ and a test question $x_t$, ICL retrieves $k$ examples related to $x_t$ from the task dataset as demonstration examples, and transforms these examples using the prompt function $f$ to form a demonstration example set $D_{k}=\{f(x_{1},y_{1}),\ldots,f(x_{k},y_{k})\}$.
The task description $I$, the example set $D_k$, and the problem is then fed into the language model for predicting $\hat{y}_{t}$. The ICL process can be expressed as follows.
\begin{equation}
  \begin{aligned}\hat{y}_{t} = \operatorname{LLM}(I,\underbrace{f(x_{1},y_{1}),\ldots,f(x_{k},y_{k})}_{\text{$k$\ demonstration examples}},\underbrace{f(x_t, \bullet)}_{\text{test\ query}})\,,
  \end{aligned}
\end{equation}
where $k$ stands for the number of demonstration examples in the ICL prompt, which typically ranges from 0 to 20.
In particular, for the special case of $k=0$, the ICL constructs the prompt without demonstration example, referred to as zero-shot learning.

\subsubsection{Chain-of-Thought (CoT) Prompting}
CoT~\cite{wei2022chain} is an improved prompting strategy that, when employed in conjunction with ICL, significantly enhances the capabilities of LLMs in tackling complex reasoning tasks.
In addition to simply constructing the prompts with a few demonstration examples as in ICL, 
CoT enriches these prompts by integrating intermediate reasoning steps, which guide the reasoning process to the final output.
Specifically, the CoT prompting strategy augments each demonstration example $\langle x, y\rangle$ in ICL with a chain-of-thought prompt $CoT$, constructing a triplet prompt $\langle x, CoT, y \rangle$.
The design of the chain-of-thought prompt $CoT$ involves both hand-crafted prompts that are independent from the problem, and the VQL sketch of each problem, which is inspired by the logical execution process of SQL queries to establish step-by-step infilling the VQL statements with the LLM.

\smallskip
\noindent\textbf{\textsc{Example 2.}}
Figure~\ref{fig_prompt_construction} presents an example of in-context learning with chain-of-thought strategy in \nltovisnospace. LLMs like ChatGPT, take the input text and infer the answer based on the task description, demonstration, and the problem. 
In the few-shot scenario, we add the most relative samples from the training dataset as examples in the demonstration. 
The demonstration part of the prompt for the \nltovis task consists of a table description, NL question, and golden VQL. 
In particular, we select the most relevant three rows of the table by calculating the Jaccard similarity correlation.

\begin{figure*}[!t]
\centering
\includegraphics[width=\textwidth]{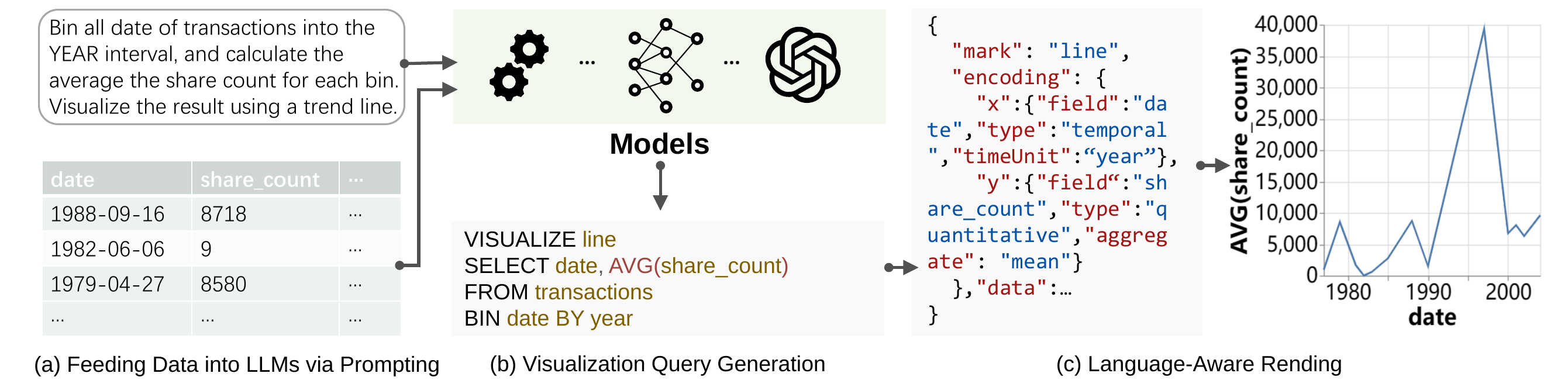}
\caption{The pipeline of \nltovisnospace. 
The task process flows from (a) feeding data into the LLMs through prompts, to (b) generating visualization queries, and finally to (c) rendering language-aware visual charts.
}
\label{fig_pipeline}
\end{figure*}

\section{LLMs for \nltovis}
\label{llms_nl2vis}
In this section, we first formulate the problem of \nltovis with a pipeline provided, and subsequently detail each individual module.

\subsection{Problem Statement}
Let $q$ denote a natural-language query, $s$ denote the schema of the table from a database to analyze. The task of \nltovis is to generate the visualization query $\hat{y}$, as follows:
\begin{equation}
    \hat{y} = \operatorname{LLM}(q, s)\,.
\end{equation}
It should be noted that the grounded tables are not restricted to one domain. 
We refer to scenarios where the databases in the test dataset have appeared in the training set as \textit{in-domain}, while databases in the test dataset are unseen as \textit{cross-domain}.

Figure~\ref{fig_pipeline} illustrates the pipeline of \nltovisnospace. The input to the models comprises a natural-language description and the grounded table. The model will then generate the visualization query in Vega-Zero. Subsequently, visualization specifications in various visual languages (e.g., Vega-Lite) can be parsed from the visualization query. 
Subsequently,
the visualization specification will be rendered into an actual visualization chart, enabling users to observe and analyze the data effectively.

\subsection{Feeding Data into LLMs via Prompting}
\label{sec: table description}
\begin{figure*}[!t]
\centering
\includegraphics[width=1\textwidth]{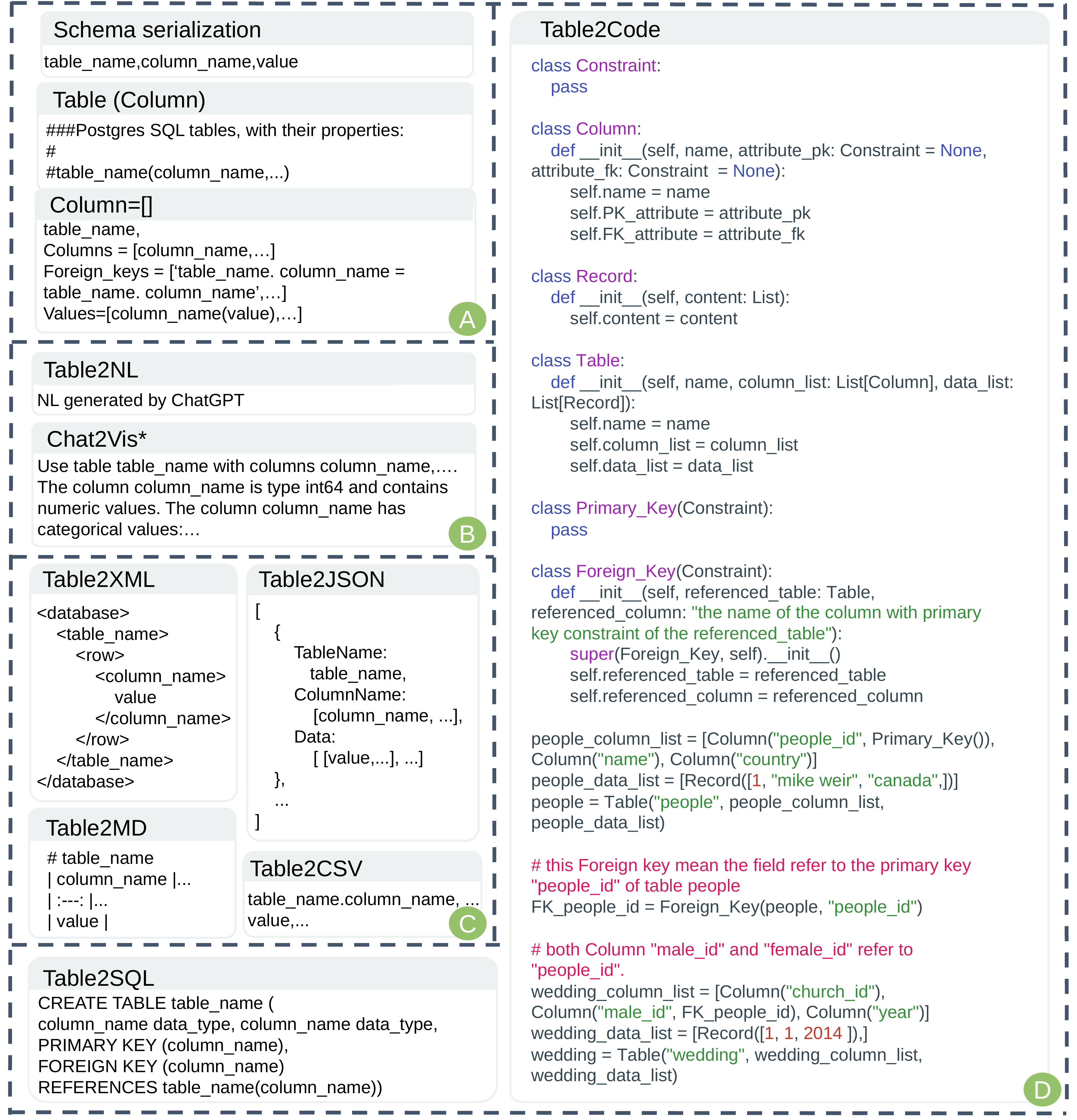}
\caption{A summary of approaches explored to transforming the table into textual prompts. (A) Table Serialization flattens database tables into linear schemas, (B) Table Summarization distills table content into concise descriptions, (C) Table Markup Formatting converts data into XML, Markdown, and CSV formats, and (D) Table Programming translates table structures into code representations.
}
\label{fig_prompt_table_description}
\end{figure*}

Given that most contemporary LLMs are primarily designed to process textual prompts due to their pre-training on sequential textual datasets, it becomes imperative to efficiently integrate structured tabular data into LLMs through effective prompting.
In this study, we investigate various strategies, i.e., table serialization, table summarization, table markup format, and table programming, to transform structured tabular data into sequential texts while preserving semantics to the greatest extent, as summarized in Figure~\ref{fig_prompt_table_description}.

\smallskip
\noindent{\textbf{A. Table Serialization} (\rounded{A} in Figure~\ref{fig_prompt_table_description}).}
Generally, previous research has proposed to feed the serialized schema into models~\cite{gong2020tablegpt}. 
For example, \textbf{Table (Column)}~\cite{liu2023comprehensive} lists each table along with its columns inside parentheses to represent the table schemas. \textbf{Column=[]} represents each table along with a list of its columns. On top of \textbf{Column=[]}, \textbf{+FK} further adds foreign keys to indicate the relationships between tables~\cite{pourreza2023din}, and \textbf{+Value} adds rows of tables. 

\smallskip
\noindent{\textbf{B. Table Summarization} (\rounded{B} in Figure~\ref{fig_prompt_table_description}).}
It is intuitive to simply describe the tables by generating a natural-language summary. Based on this intuition, Chat2Vis~\cite{maddigan2023chat2vis} uses a description built from a template to transform the table into a text prompt. 
The description is comprised of individual entries, each signifying the data type of a corresponding column. 
In this paper, we generate the table summaries by invoking the API of ChatGPT.
Specifically, we accomplish this by flatting the table column names and values into a prompt, delineated as ``\texttt{[TABLE]}, \textit{Describe the tabular data in text, including all metadata such as its name and type.}''.

\smallskip
\noindent{\textbf{C. Table Markup Formating} (\rounded{C} in Figure \ref{fig_prompt_table_description}).}
In this strategy, we explore describing the tabular data using markup languages, including CSV, JSON, Markdown, and XML.
\textbf{Table2CSV} converts tabular data into CSV format, where each line represents a data record containing one or more comma-separated fields.
\textbf{Table2JSON} aims to convert tabular data into a JSON object, where data is represented in <name, value> pairs, enclosed by curly braces ``\{\}'' for objects and square brackets ``[]'' for arrays.
\textbf{Table2MD} aims to convert the table into a Markdown file, which uses simple and intuitive syntax to denote structure and style.
\textbf{Table2XML} aims to convert the table into an XML file, which is an eXtensible Markup Language used to represent structured data in a human-readable and machine-readable way.

\smallskip
\noindent{\textbf{D. Table Programming} (\rounded{D} in Figure~\ref{fig_prompt_table_description}).}
In order to feed the structured tabular data into LLMs, we propose to represent the structured tabular data in programming languages. 
\textbf{Table2SQL}
uses \texttt{Create} statements to describe database schema~\cite{raffel2020exploring}, which emphasizes the relationship among tables.
For showcasing the content of a database, \textbf{+Select}~\cite{raffel2020exploring} employs the ``\texttt{SELECT * FROM Table LIMIT R}'' query to display the first $R$ rows of each table.
\textbf{Table2Code} proposes to transform the tabular data into general-purpose programming languages. In particular, 
we resort to the Python programming language, which effectively 
employs an inherent object-oriented paradigm and is well-suited for encoding structured data~\cite{wang2022code4struct}. 
Moreover, we utilize Python's type hinting feature and represent tabular data using a class-based representation in Python. 
We define classes of \texttt{Table}, \texttt{Column}, \texttt{Constraint}, \texttt{Primary\_Key}, \texttt{Foreign\_Key}, and \texttt{Record}, followed by the instantiation of these classes to create a table object.

\subsection{Visualization Query Generation}
We utilize APIs provided by OpenAI in the GPT-3.5 and GPT-4 series to engage with LLMs. 
Specifically, we initially represent the tabular data with the format as outlined in Figure \ref{fig_prompt_table_description}. Then, we construct the tables and queries into prompts as delineated in Figure \ref{fig_prompt_construction}. Finally, these prepared prompts are subsequently fed into LLMs via API calls.
Our research methodology can be applied to any LLMs that support prompting.

\subsection{Language-Aware Rending}
According to nvBench~\cite{luo2021nvbench}, the visualization query can be presented as the tree format as introduced in~\cite{luo2021synthesizing} for fair evaluation of grammar. These \textit{Abstract Syntax Trees (ASTs)} can then be translated to target visualization specification syntax, so as to render the visualization charts. 
The translation from a visualization query to a target visualization specification is hard-coded based on the grammar of ASTs. 
Currently, the nvBench provides a module of Python 3 code~\cite{luo2021synthesizing} for converting visualization query to Vega-Lite.
Thus, the pipeline is fully automated and enables an effortless evaluation of the visualization query.

\section{Evaluation Setup}
\label{evaluation_setup}
We investigate the capability of LLMs in automating data visualization from natural-language descriptions, by answering the following three \textit{Research Questions (RQs)}.

\begin{itemize}
    \item \textbf{RQ1 [Prompt Engineering]:} How to feed the natural-language query as well as the structural table into LLMs via prompting?
    \item \textbf{RQ2 [Overall Performance]:} How do the LLMs perform when compared with several existing models in \nltovisnospace? 
    \item \textbf{RQ3 [Iterative Updating]:} Can the results be iteratively updated via some optimization strategies?
\end{itemize}
To address RQ1, we first explore the conversion of structured tables into sequential prompts, followed by an examination of how the content of these tables influences the results.
To answer RQ2, we conduct a comparative analysis of the performance exhibited by both conventional neural networks and LLMs concerning \nltovisnospace.
Furthermore, we analyze the impact of the number of in-context demonstrations on the performance of LLMs.
To answer RQ3, we delve into instances of failure within the LLMs and subsequently propose strategies for enhancing model performance through optimization techniques.

All experiments are conducted on a machine with 252 GB memory and 4 Tesla V100 32GB GPUs. We use the default hyperparameter settings provided by each method. Besides, we split the dataset into a training dataset, a valid dataset, and a test dataset with 7:2:1 based on in-domain and cross-domain settings, as explained in Sec.~\ref{dataset}. We use the test dataset for evaluation and the training dataset for demonstrations.

\subsection{Dataset}
\label{dataset}
We employ the widely-recognized benchmark nvBench~\cite{luo2021nvbench} to assess performance in the \nltovis task.
The nvBench dataset encompasses 780 relational tables sourced from 153 databases across 105 diverse domains such as sports, colleges, hospitals, and more. It features 7,247 visualizations, resulting in 25,750 pairs of natural-language descriptions and corresponding visualizations.

\noindentparagraph{\textbf{\textup{Domain Setting.}}}
In prior studies~\cite{luo2021synthesizing,luo2021natural}, datasets are predominantly partitioned randomly based on visualization queries, without considering database divisions. Upon re-implementing the ncNet\footnote{\url{https://github.com/Thanksyy/ncNet}}, we notice that databases from the test set are exposed during the training process, constituting an in-domain setting. 
We propose a cross-domain setting by partitioning the dataset such that there is no overlap between databases in the training and test datasets. In this setup, models are required to predict queries on the unseen databases.     
To ensure a fair and comprehensive evaluation, we conduct experiments in both in-domain and cross-domain settings.

\noindentparagraph{\textbf{\textup{Multi-Table Setting.}}} 
In our experiments, we categorize scenarios based on the number of tables involved in the input. 
Instances with multiple tables are designated as ``\textit{join}'' scenarios, while those centered on a single table are categorized as ``\textit{non-join}'' scenarios.
In ``\textit{join}'' scenarios, our objective entails generating visualizations by integrating information from multiple tables. This process involves merging data based on shared columns, presenting challenges to both data integration and the subsequent visualization efforts.
Conversely, ``\textit{non-join}'' cases involve generating visualizations from individual tables, acting as a benchmark for evaluating the model's ability to manage less complex data structures.

\subsection{Evaluation Metrics}
\label{evaluation}
\begin{figure}[!t]
\centering
    \begin{subfigure}[b][][c]{.31\textwidth}
	\centering
        \includegraphics[width=\textwidth]{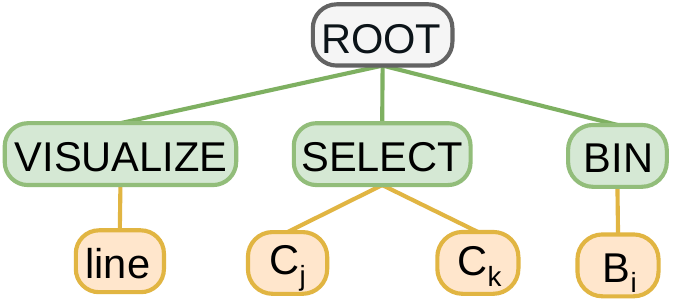}
        \caption{$AST_1$}
        \label{fig_ast1}
  \end{subfigure}
  \hspace{3em}
  \begin{subfigure}[b][][c]{.31\textwidth}
	\centering
        \includegraphics[width=\textwidth]{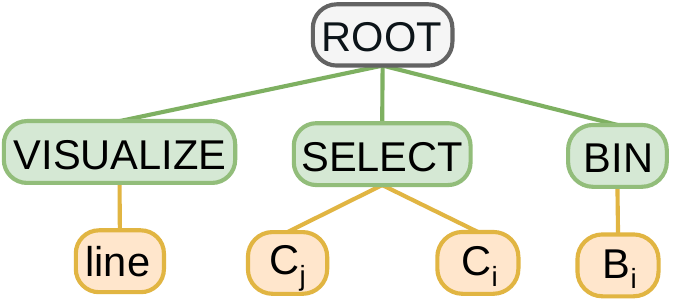}
        \caption{$AST_2$}
        \label{fig_ast2}
  \end{subfigure}
  \caption{An illustration of ASTs of VQL queries. Even though they may not be exactly matched, their execution results are identical.
}
\label{fig_ast}
\end{figure}

To ensure a thorough and equitable assessment of the generated VQL queries, we employ three widely recognized metrics: exact accuracy, execution accuracy, and component accuracy, as established in the \nltovis task~\cite{luo2021synthesizing}.
Before introducing these metrics, we first introduce two visualization queries represented as ASTs, as shown in Figure~\ref{fig_ast}.
Each of them has three subtrees based on the grammar of VQL described in Sec.~\ref{vql}. Specifically, the node of \texttt{VISUALIZE} denotes a line chart type. The \texttt{SELECT} node combines the attributes of the columns $C_j$ and $C_k$. The \texttt{BIN} node with $B_i$ sets a bucket of values in the temporal column.

\noindentparagraph{\textbf{\textup{Exact Accuracy~\cite{luo2021synthesizing}.}}}
This metric is designed to assess the exact match between the predicted AST and the ground-truth AST of VQL queries. 
It can be formulated as 
$Acc_{AST} = N_{AST} / N$, where $N_{AST}$ represents the count of generated ASTs that are exactly equivalent
to the ground truth ASTs in each node, and $N$ signifies the total number of ASTs under consideration. 

\noindentparagraph{\textbf{\textup{Execution Accuracy~\cite{luo2021synthesizing}.}}}
This metric measures the accuracy of the visualization results by determining whether the predicted visualization aligns with the ground truth. It can be calculated using the formula $Acc_{exe} = N_{exe} / N$, where $N_{exe}$ denotes the number of VQL whose results match the ground truth in the execution, and $N$ denotes the total number of visualizations.
From Figure~\ref{fig_ast}, it is apparent that the subtrees rooted at ``\texttt{VISUALIZE}'' and ``\texttt{BIN}'' retain consistent structures between $AST_{1}$ and $AST_{2}$. While variations within the ``\texttt{SELECT}'' subtrees suggest they are not exact matches.
Even though, the VQL execution results could be matched if $x/y/z$-axis data (executed by VQL) are correct, as exemplified by queries such as ``\texttt{VISUALIZE line SELECT date, COUNT(date) from payments BIN date by month}'' and ``\texttt{VISUALIZE line SELECT date, date\_count from payments BIN date by month}''.

\subsection{Comparison Models for \nltovis}
\label{baselines}

\noindentparagraph{\textbf{\textup{Traditional Neural Models.}}}

\underline{\textsc{Seq2Vis}}~\cite{luo2021synthesizing} is a sequence-to-sequence model that initially encodes the natural-language query into a hidden embedding using an LSTM and subsequently decodes it into a visualization through another LSTM.
\underline{Transformer}~\cite{vaswani2017attention} is another encoder-decoder network that has been considered a foundational component in LLMs. We also apply the Transformer into \nltovisnospace.
Based on Transformer, \underline{ncNet}~\cite{luo2021natural} designs several novel visualization-aware optimizations, such as using attention-forcing to optimize the learning process and visualization-aware rendering to produce better visualization results. 
\underline{RGVisNet}~\cite{song2022rgvisnet} 
is a hybrid framework for \nltovis to retrieve, refine, and generate visualizations from a large codebase using a graph neural network.

\noindentparagraph{\textbf{\textup{LLMs (Finetuned Models).}}}
T5~\cite{shaw2020compositional} is an encoder-decoder model pre-trained on a multi-task mixture of unsupervised and supervised tasks for which each task is converted into a text-to-text format. 
\underline{T5-Small} is a released T5 model with 60 million parameters. 
\underline{T5-Base} is a released T5 model with 220 million parameters. 

\noindentparagraph{\textbf{\textup{LLMs (Inference-only Models).}}}
\underline{Chat2Vis}~\cite{maddigan2023chat2vis} leverages the power of \texttt{code-davinci-002} to enable users to generate data visualizations using natural-language queries in Python plots and measures results against nvBench. 
Our evaluation focuses on the models accessible via the OpenAI API: GPT-3.5 (\texttt{text-davinci-002}, \texttt{text-davinci-003}, \texttt{gpt-3.5-turbo-16k}), and GPT-4 (\texttt{gpt-4}). 
GPT-3.5 is a set of models built on the InstructGPT~\cite{ouyang2022training}, trained on a large corpus of programming languages and natural languages.
\underline{\texttt{text-davinci-003}}\footnotemark
~is fine-tuned by reinforcement learning from human feedback~\cite{NIPS2017_d5e2c0ad}, which improves its ability to generate better quality and longer output.
\underline{\texttt{text-davinci-002}}\footnotemark[\value{footnote}]\footnotetext{
Note that this study was conducted between April and October 2023. Subsequently, as of January 2024, the \texttt{text-davinci-003} and \texttt{text-davinci-002} have been upgraded to \texttt{gpt-3.5-turbo-instruct}. More details are referred to OpenAI documentation: \url{https://platform.openai.com/docs/deprecations}.
} is trained with supervised fine-tuning instead of reinforcement learning.
\underline{\texttt{gpt-3.5-turbo-16k}} is
optimized for chat applications and increased input length.
GPT-4 is a large multimodal model that is capable of solving complex problems with greater accuracy than any of the previous models, due to its excellent general knowledge and reasoning capabilities~\cite{openai2023gpt4}.
\underline{\texttt{gpt-4}} 
revolutionizes loss function computation by incorporating a scaling law and an irreducible loss term~\cite{openai2023gpt4blog}, enabling precise prediction of final loss within the internal codebase.

\section{Results and Analysis}
\label{results}
In this section, we present the experimental results for each RQ with comprehensive analysis.
\subsection{RQ1: Prompt Engineering}
\label{RQ1_result}

In this RQ, we investigate various strategies for inputting structured tabular data into LLMs through prompting and analyze which aspect of the table content has the greatest influence on \nltovisnospace.
\subsubsection{RQ1-1: How to transform the structured tabular data into sequential prompts?}
\label{RQ1-1}

We evaluate the performance of the model \texttt{text-davinci-003} while varying the methods for transforming tabular data into textual prompts. 
We carry out this assessment in both cross-domain and in-domain settings. Additionally, we investigate configurations for both non-join and join scenarios, where single-table and multi-table contexts are considered, respectively.
Note that when transforming tabular data into a markup format, we only consider a single row of content, specifically the one most relevant to the input question, as determined by Jaccard similarity.

\begin{table*}[t!]
    \centering
    \caption{Performance of the model \texttt{text-davinci-003} while varying the methods for transforming tabular data into textual prompts, under both the cross-domain and in-domain settings.
    }
    \setlength{\tabcolsep}{4pt} 
    \begin{tabular}{l|cc|cc|cc|cc|cc|cc}
        \hline
        &\multicolumn{6}{c|}{\textbf{Cross-domain}} &\multicolumn{6}{c}{\textbf{In-domain}} \\
        \cline{2-13}
        &\multicolumn{2}{c|}{\textbf{Non-join}} & \multicolumn{2}{c|}{\textbf{Join}} & \multicolumn{2}{c|}{\textbf{Overall}}&\multicolumn{2}{c|}{\textbf{Non-join}} & \multicolumn{2}{c|}{\textbf{Join}} & \multicolumn{2}{c}{\textbf{Overall}}\\
        &\multicolumn{1}{c}{\textbf{Exa.}} & \multicolumn{1}{c|}{\textbf{Exe.}} 
        &\multicolumn{1}{c}{\textbf{Exa.}} & \multicolumn{1}{c|}{\textbf{Exe.}}
        &\multicolumn{1}{c}{\textbf{Exa.}} & \multicolumn{1}{c|}{\textbf{Exe.}}
        &\multicolumn{1}{c}{\textbf{Exa.}} & \multicolumn{1}{c|}{\textbf{Exe.}} 
        &\multicolumn{1}{c}{\textbf{Exa.}} & \multicolumn{1}{c|}{\textbf{Exe.}}
        &\multicolumn{1}{c}{\textbf{Exa.}} & \multicolumn{1}{c}{\textbf{Exe.}}\\
        \hline
        \multicolumn{13}{l}{\textbf{\textit{Table Serialization}}}\\
        \hline
        Schema &0.33	&0.50	&0.06 &0.09 &0.32 &0.47 & 0.30 &0.31 & 0.27 & 0.23 &0.29 & 0.28\\
        Table (Column)	&0.37 &0.56	&0.09 &0.11 &0.36 &0.54 &0.60 &0.59 &0.52 &0.44 &0.57 &0.53\\ 
        Column=[]	&0.37	&0.54	&0.11 &0.12 &0.36 &0.52 &0.61 &0.61 &0.52 &0.45 &0.58 &0.55\\ 
        \hline
        \multicolumn{13}{l}{\textbf{\textit{Table Summarization}}}\\
        \hline
        Table2NL	&0.35	&0.54	&0.13 &0.17 &0.34 &0.52 &0.63 &0.63 &0.53 &0.46 &0.60 &0.57\\
        Chat2Vis*	&0.37	&0.57	&0.14 &0.20 &\textbf{0.36} &\textbf{0.55} &0.58 &0.61 &0.32 &0.28 &0.49 &0.49\\ 
        \hline
        \multicolumn{13}{l}{\textbf{\textit{Table Markup Format}}}\\
        \hline
        Table2JSON  &0.36	&0.54	&0.08 &0.15 &0.35 &0.52 &0.63 &0.63 &\textbf{0.56} &\textbf{0.48} &\textbf{0.61} &0.57\\
        Table2CSV   &0.35	&0.52	 &0.07 &0.12 &0.34 &0.50 &0.59 &0.59 &0.54 &0.46 &0.57 &0.54\\ 
        Table2MD   &0.36	&0.53	 &0.07 &0.09 &0.34 &0.50 &0.59 &0.61 &0.53 &0.46 &0.57 &0.56\\ 
        Table2XML   &0.36	&0.54	&0.08 &0.16 &0.35 &0.52 &0.63 &0.64 &0.56 &0.47 &\textbf{0.61} &\textbf{0.58}\\ 
        \hline
        \multicolumn{13}{l}{\textbf{\textit{Table Programming}}}\\
        \hline
        Table2SQL   &\textbf{0.39}	&\textbf{0.58}	&\textbf{0.17} &\textbf{0.25}  &\textbf{0.38} &\textbf{0.55} &\textbf{0.64} &\textbf{0.64} &0.53 &0.47 &\textbf{0.61} &\textbf{0.58}\\ 
        \textbf{Table2Code}	&0.36 &\textbf{0.59}	&\textbf{0.18}	&0.08  &0.34 &\textbf{0.56} &0.53 &0.58 &0.47 &0.43 &0.55 &0.55\\ 
        \hline
    \end{tabular}
    \label{tab_table_description_GPT3}
\end{table*}

Table~\ref{tab_table_description_GPT3} presents the model performance for \texttt{text-davinci-003} across various methods of transforming tabular data into textual prompts, considering both cross-domain and in-domain scenarios. All results are derived from a single-shot example provided in the ICL.
From this table, it is clear that feeding LLMs with tables by encoding them in programming languages is the most effective method, both in the in-domain and cross-domain settings. 
Specifically, under the in-domain setting, prompting table via \textit{Table2SQL}, \textit{Table2XML}, and \textit{Table2JSON} achieves the best performance, reaching up to 61\% in terms of the Exact Accuracy.
While under the cross-domain setting, prompting table via \textit{Table2Code}, \textit{Table2SQL}, and \textit{Chat2Vis*} achieves the best Execution Accuracy of 56\%, 55\%, and 55\%, respectively.
It suggests that converting structured tabular data into machine-readable markup formats and using general-purpose programming languages can effectively 
design prompts to interact with LLMs for \nltovisnospace.
When comparing the prompts of \textit{Table2Code} to \textit{Chat2Vis*}, we can see that \textit{Table2Code} obtains 6\% and 1\% improvement in terms of the Execution Accuracy in the in-domain and cross-domain settings, respectively.
Interestingly, we observe that the LLM-based model exhibits slight fluctuations in the performance of several settings.
In cross-domain scenarios, the optimal performance is achieved through the application of table programming to transform input tables; however, this trend does not persist in in-domain settings.
One plausible explanation for this observation is that, in the in-domain scenario, where the data adheres to a similar distribution, the model has the opportunity to discern the schema of the test database from demonstration examples during in-context learning. Consequently, the impact of exploring table transformations becomes diminished within this specific in-domain context.

Furthermore, when comes to the non-join and join scenarios under the cross-domain setting, we can see that our proposed methods of converting tabular data into code can still maintain the best performance.
For example, \textit{Table2Code} surpasses the state-of-the-art \textit{Chat2Vis*} by 2\% and 4\% in terms of the Execution Accuracy and Exact Accuracy when handling the non-join cases and join cases, respectively.
We attribute the improvements to the effective preservation of inner structural information within tables when concerting structured tabular data into source code. This benefit arises from the inherent interplay between the structural aspects of source code and tabular data.

\begin{tcolorbox}
\textbf{Finding 1-1.} 
Converting structured tabular data into machine-readable markup formats (e.g., JSON, XML) or utilizing general-purpose programming languages (e.g., SQL, Python) yields superior performance compared to relying solely on natural-language summaries or straightforward table serialization.
\end{tcolorbox}
\begin{figure}[!t]
\centering
    \begin{subfigure}[b][][c]{.44\textwidth}
	\centering
        \includegraphics[width=\linewidth]{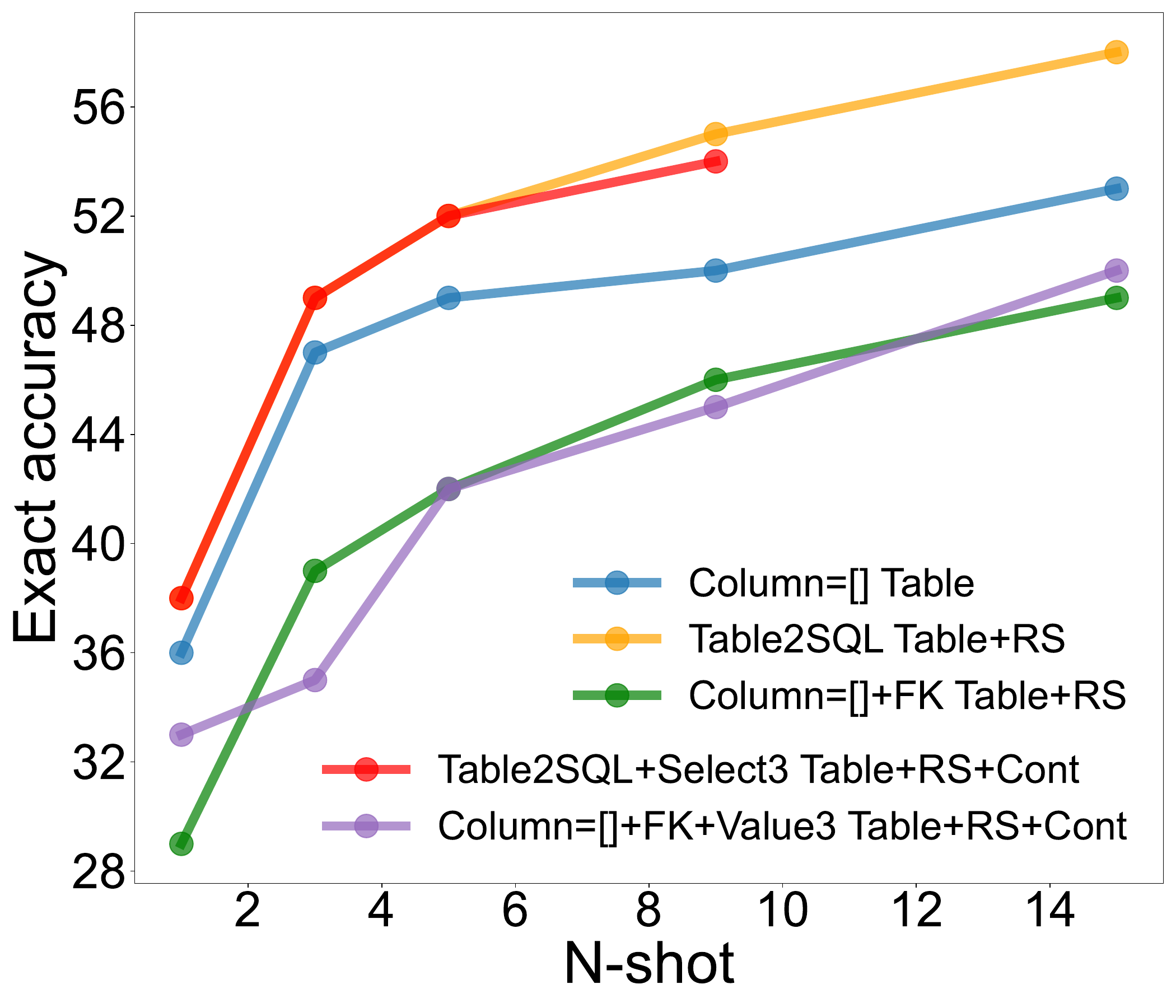}
	\caption{ Exa. Acc. in cross-domain}
	\label{fig_content_cross_exact}
    \end{subfigure}
    \begin{subfigure}[b][][c]{.44\textwidth}
	\centering
        \includegraphics[width=\linewidth]{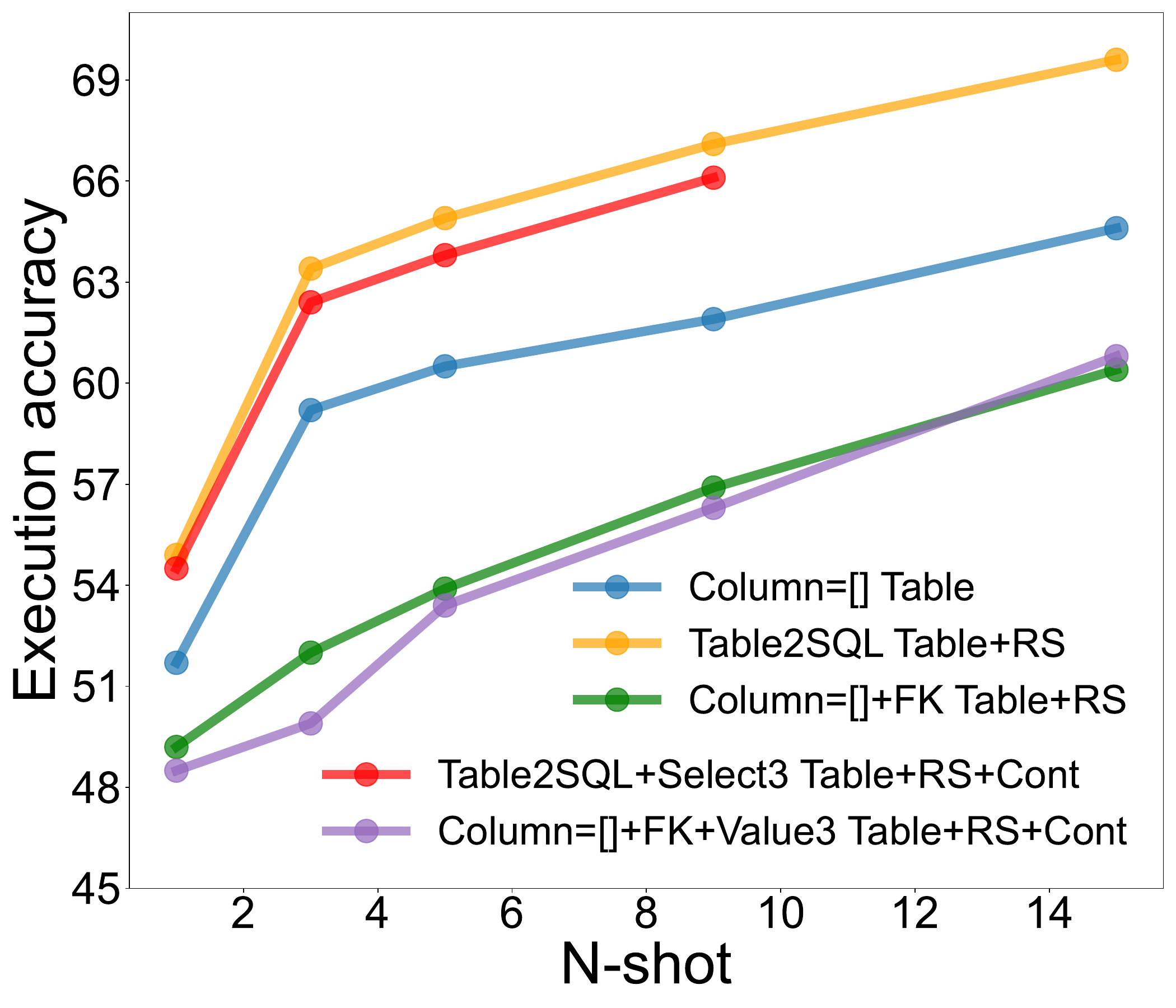}
	\caption{ Exe. Acc. in cross-domain}
	\label{fig_content_cross_exec}
    \end{subfigure}
    \begin{subfigure}[b][][c]{.44\textwidth}
	\centering
        \includegraphics[width=\linewidth]{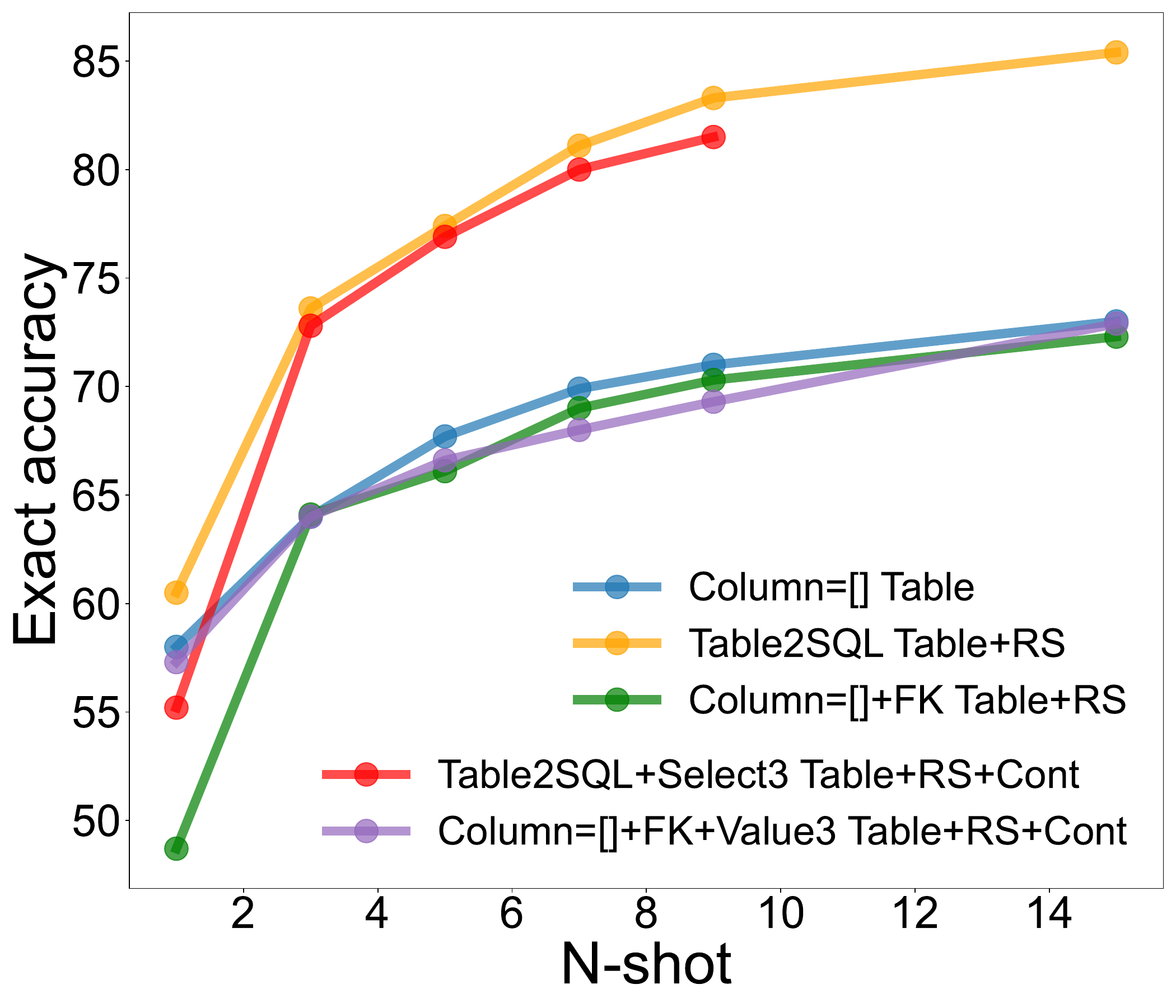}
	\caption{ Exa. Acc. in in-domain}
	\label{fig_content_in_exact}
    \end{subfigure}
    \begin{subfigure}[b][][c]{.44\textwidth}
	\centering
        \includegraphics[width=\linewidth]{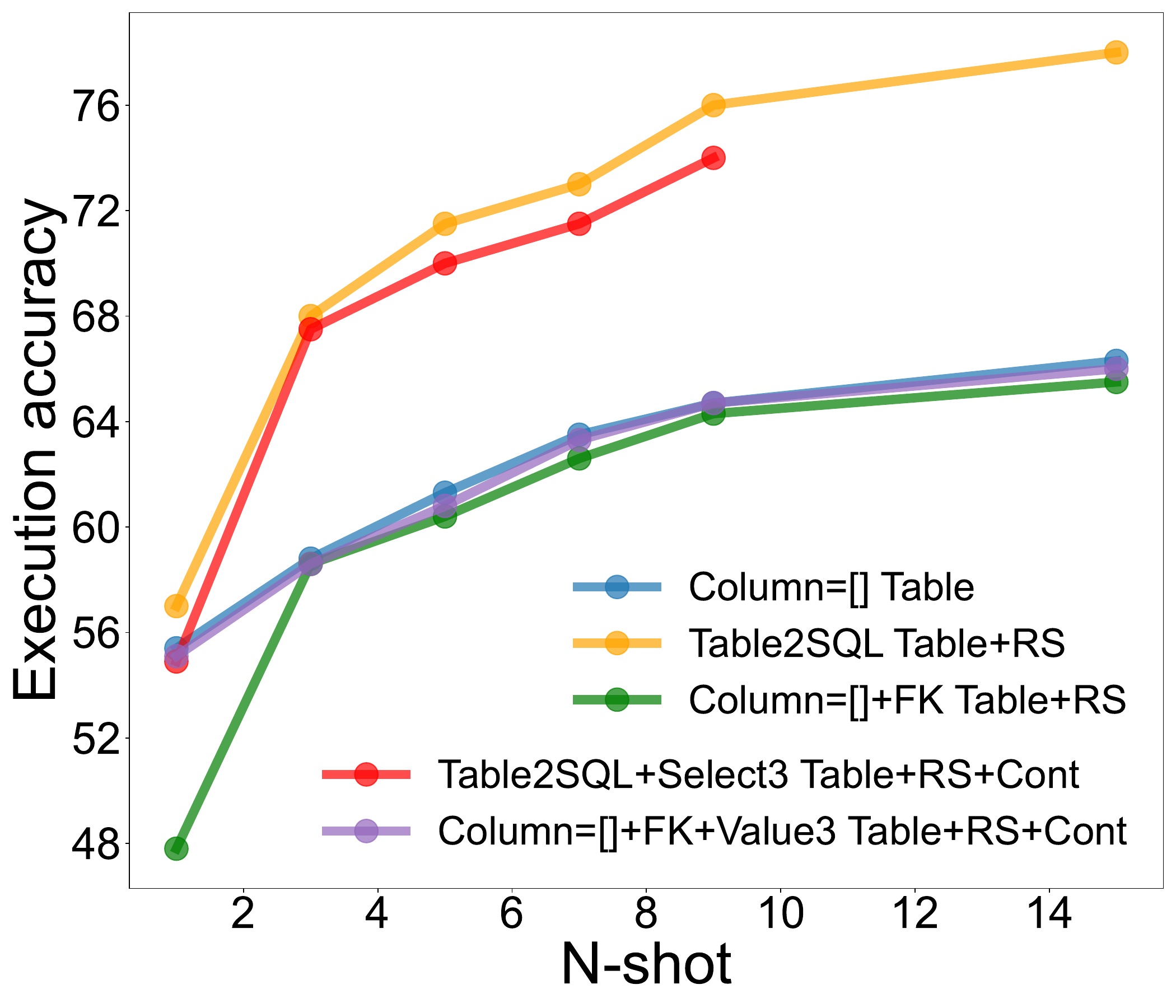}
	\caption{ Exe. Acc. in in-domain}
	\label{fig_content_in_exec}
    \end{subfigure}
\caption{Exact Accuracy and Execution Accuracy of \texttt{text-davinci-003} under the cross-domain and in-domain settings, with 1, 3, 5, 7, and 15 demonstrations provided in in-context learning. RS and Cont correspond to the table relationship and table content, respectively.
}
\label{fig_content}
\end{figure}

\subsubsection{RQ1-2: What table content should be considered in \nltovisnospace?}
\label{RQ1-2}
We further dive into analyzing the importance of table components in prompt construction and pinpoint the most reliant table component in different domain and few-shot settings. 
We categorize table contents into three levels:
(i) \textit{Table Schema}, which encompasses table names and column names,
(ii) \textit{+RS}, denoting relationships within the table, such as foreign keys,
and (iii) \textit{+Cont}, encompassing the actual content of the table, including the data in table row values.
To simplify our experiment, we have chosen table serialization as the primary prompting approach. More specifically, our exploration encompasses three distinct variants:
(i) \textit{Column=[]}, 
(ii) \textit{Column=[]+FK}, \textit{Table2SQL}, and 
(iii) \textit{Column=[]+FK+Value3}, \textit{Table2SQL+Select3}.


Figure~\ref{fig_content} shows the Exact Accuracy and Execution Accuracy of \texttt{text-davinci-003} under the cross-domain and in-domain settings, with few demonstrations provided in in-context learning.
From this figure, it is evident that the performance of the \textit{Table Column=[]} configuration surpasses that of the \textit{Table+RS Column=[]+FK} and \textit{Table+RS+Cont Column=[]+FK+Value} configurations in both in-domain and cross-domain settings. This observation underscores the significance of the table schema as a comprehensive element within the prompt for \textit{Column=[]}.

Moreover, when comparing the disparity between \textit{Table Column=[]} and \textit{Table+RS Column=[]+FK}, we observe that in the in-domain setting, the gap slightly narrows as the number of examples increases. This suggests that table knowledge in prompts has minimal influence on input, likely due to the fact that, in the in-domain setting, LLMs have seen test tables during demonstrations from the same database domain. However, a significant gap emerges in the cross-domain setting (i.e., increasing from 2.5\% to 7.2\% in terms of Execution Accuracy as the number of examples increases from 1 to 3), indicating that the additional table knowledge, specifically table relationships and content, in the \textit{Column=[]} prompt is superfluous. Moreover, this redundancy could potentially introduce complexity and adversely impact the performance of LLMs in the \nltovis task.

Additionally, when comparing the performance of \textit{Table+RS Table2SQL} and \textit{Table+RS+Cont Table2SQL+Select3}, we observe that there is no reduction in the gap between these two prompts. This observation suggests that \textit{Table2SQL} remains effective even without the inclusion of row values. This can be attributed to the nature of the \nltovis task, where the importance of table content often takes a secondary role. Users typically only need to reference table and column names to convey their visualization requirements. Essentially, the model's primary task is to establish a seamless connection between the natural-language query and the database schema, subsequently generating the visualization query successfully. This underscores that table content may not play a significant role in this context.

Finally, when comparing the performance of \textit{Table+RS Column=[] +FK} and \textit{Table+RS Table2SQL}, we can observe that while both prompts incorporate table schema and table relationship knowledge, they yield markedly distinct results. This observation underscores the substantial impact of well-crafted formats for representing structured tabular data.

\begin{tcolorbox}
\textbf{Finding 1-2.} 
Table schema plays an important role in the task of \nltovisnospace, both under the cross-domain and in-domain settings. 
\end{tcolorbox}

\begin{table}[t!]
        \centering
        \caption{Results of comparing LLMs with several baselines. 
        }
    \begin{tabular}{l|cc|cc}
        \hline
        &\multicolumn{2}{c|}{\textbf{Cross-domain}} & \multicolumn{2}{c}{\textbf{In-domain}} \\
        \cline{2-5}
        &\multicolumn{1}{c|}{\textbf{Exa. Acc.}} & \multicolumn{1}{c|}{\textbf{Exe. Acc.}} &\multicolumn{1}{c|}{\textbf{Exa. Acc.}} & \multicolumn{1}{c}{\textbf{Exe. Acc.}}\\
        \hline
        \multicolumn{5}{l}{\textbf{\textit{Baseline}}}\\
        \hline
        \textsc{Seq2Vis}	&0.02		&-	  &0.66		&- 	 \\
        Transformer	&0.03		&-	 &0.73		&- 	 \\
        ncNet	&0.26		&-	  &\textbf{0.77}		&-	 \\
        RGVisNet	&\textbf{0.45}		&-	 &-		&- 	 \\
        Chat2Vis	&-		&0.43	  &-		&- 	\\            
        \hline
        \multicolumn{5}{l}{\textbf{\textit{Finetuned}}}\\
        \hline
        T5-Small &0.60		&0.61 	&0.92  &0.81    \\
        T5-Base &0.71		&0.72 	&0.93  &0.82    \\
        \hline
        \multicolumn{5}{l}{\textbf{\textit{Inference-only}}}\\
        \hline
        \texttt{text-davinci-002}	&0.57		&0.68	 &0.84	&0.75 	\\
        \texttt{text-davinci-003}	&0.58		&0.70	 &0.87	&0.77 	\\
        \texttt{gpt-3.5-turbo-16k} &0.56	&0.63  &0.59	&0.59 \\
        \texttt{gpt-4}	&0.61	&0.72   &0.83		&0.74 \\
        \hline
    \end{tabular}
    \label{tab_LLMs_Performance}
\end{table}

\begin{figure}[!t]
\centering
    \begin{subfigure}[b][][c]{.44\textwidth}
	\centering
        \includegraphics[width=\linewidth]{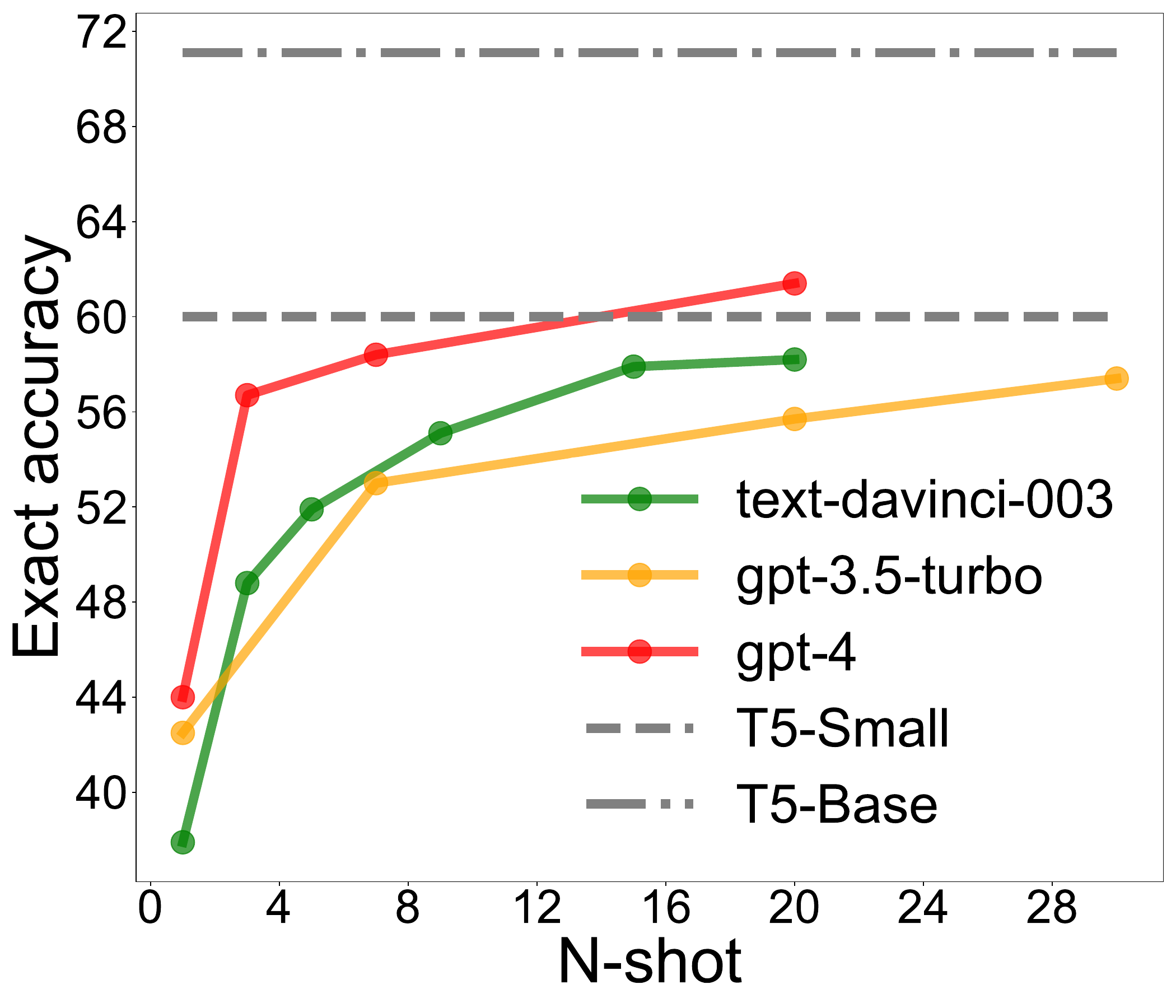}
	\caption{Exa. Acc.}
	\label{fig_trend_a}
    \end{subfigure}
    \begin{subfigure}[b][][c]{.44\textwidth}
	\centering
        \includegraphics[width=\linewidth]{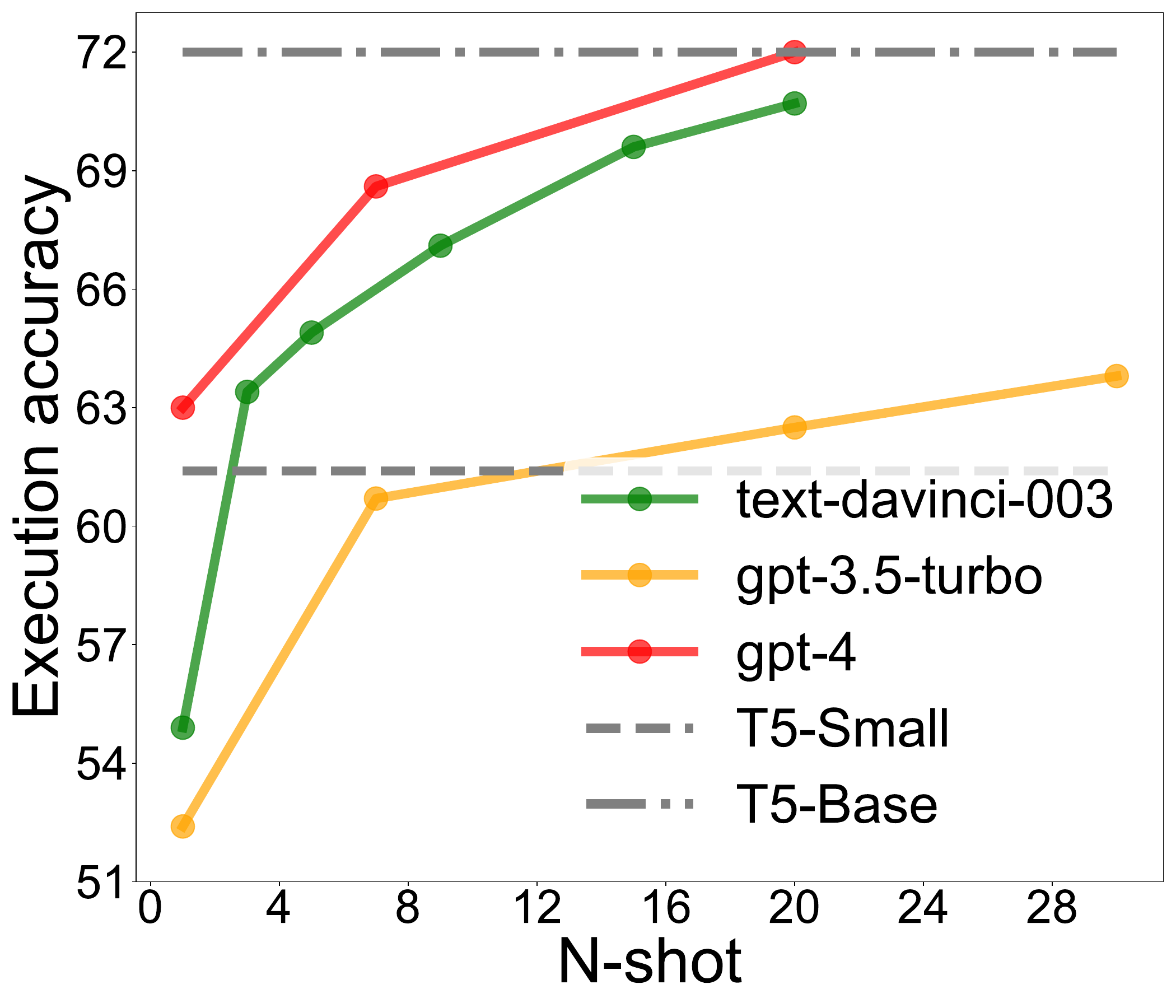}
	\caption{Exe. Acc.}
	\label{fig_trend_b}
    \end{subfigure}
\caption{Exact Accuracy and Execution Accuracy with varying number of support examples. The x-axis indicates the number of few-shot examples used.}
\label{fig_finetune}
\end{figure}

\subsection{RQ2: Overall Performance}
\label{RQ2_result}

In this RQ, we investigate the overall performance of LLMs, and compare them with several traditional neural models for \nltovisnospace, in both in-domain and cross-domain settings.

\subsubsection{RQ2-1: How do the LLMs perform when compared with existing works?}
\label{RQ2-1}
To answer this RQ, we conduct a comparative analysis to evaluate LLMs, including the fine-tuned T5 models together with the inference-only GPT-3.5 and GPT-4 series models, against several state-of-the-art baseline models.
We fine-tune T5-Small and T5-Base for the \nltovis task with a maximum of 11.6$k$ steps and 100$k$ steps, respectively. 
Our configurations specify a maximum learning rate of $1 \times 10^{-3}$, a batch size of 4, and a dropout rate of 0.1.
To set up a fair comparison, we explore inference-only models using the same \textit{Table2SQL} prompt with 20 examples provided in the ICL (20-shot).

Table~\ref{tab_LLMs_Performance} shows the overall performance of LLMs as well as several baselines, under both the cross-domain and in-domain settings for the \nltovis task.
From this table, it is clear that LLMs (both the finetuned models and inference-only models) significantly outperform the traditional baselines, both in the in-domain and cross-domain settings. 
A closer examination of in-domain performance reveals that the predominant baseline models, namely, \textsc{Seq2Vis}, Transformer, and ncNet, achieve their peak scores at 66\%, 73\%, and 77\%, respectively. This underscores the role of ncNet in enhancing the seq2seq models through attention forcing.
Furthermore, the fine-tuned language models, T5-Small and T5-Base, significantly outperform state-of-the-art methods with scores of 92\% and 93\%, respectively.
For in-context learning, inference-only models also achieve commendable scores. Notably, \texttt{text-davinci-003} stands out with an 87\% score, surpassing existing baseline methods by 10\%.
This underscores the idea that LLMs, trained on natural-language text or code corpora, represent the preeminent models for the task. The enhancement is attributed to the robust capabilities of LLMs in comprehending the inherent knowledge, including table schema and structure, within the same domain, leading to superior responses to new queries.

Furthermore, when comparing performance in cross-domain scenarios, we are genuinely surprised by the substantial decline observed across all baseline models (e.g., ncNet, plummeting from 77\% to 26\%). It is noteworthy that RGVisNet is the first to address this phenomenon and succeeds in boosting cross-domain performance by 45\%. This underscores the vast potential for enhancement in this context. The underperformance of traditional neural models, designed for random-splitting data settings, becomes evident as they struggle to generalize to previously unseen databases and grapple with the task of linking natural language to table schema.
However, the fine-tuned models, specifically T5-Small (60\%) and T5-Base (71\%), significantly outshine the baseline models. Among the inference-only models, \texttt{gpt-4} shows the most impressive improvement, surging by 61\%, while \texttt{gpt-3.5-turbo-16k} lags behind with a 56\% increase. This serves as a testament to the exceptional generalization capabilities of LLMs, empowering them to effectively assimilate new knowledge and establish connections with queries when applied to previously uncharted databases. The result is an unequivocal improvement in visualization results.

Lastly, in comparing the performance of fine-tuned models with that of inference-only models, we observe that while fine-tuned models exhibit the most desirable performance, their advantages in fine-tuning can be equaled by inference-only models with in-context learning. 
As shown in Figure~\ref{fig_finetune}, both fine-tuned T5-Small and T5-Base models achieve execution accuracies of 61\% and 72\%, respectively. 
The models \texttt{text-davinci-003} and \texttt{gpt-3.5-turbo-16k} demonstrate superior performance compared to the T5-Small model, particularly with 3-shot and 13-shot scenarios. This suggests that, after exposure to an ample number of examples in contextual learning, LLMs acquire the capability to utilize demonstrations for precise inference and the generation of visualizations in tables not encountered before.
Additionally, we investigate the comparative costs of fine-tuned models and inference-only models with 20-shot in-context learning. 
From Table~\ref{tab_model_compare}, we can observe that even though the inference-only models typically incur significant computational costs due to their huge size, they offer considerable time savings through in-context learning.

\begin{table}[t!]
    \centering
    \caption{Statistics of model parameters and cost time.}
    \begin{tabular}{l|c|c|c}
        \hline
        & \textbf{Parameters}& \textbf{Cost Time} & \textbf{Model Size} \\
        \hline
        T5-Small           & 60M & 3 days & 200MB             \\
        T5-Base            & 220M & 5 days & 500MB             \\
        \texttt{text-davinci-003}   & 1.5B & 15 hours & 1GB             \\
        \texttt{gpt-3.5-turbo-16k}  & 4B & 4 hours & 2GB             \\
        \texttt{gpt-4}              & - & 4 hours & -            \\
        \hline
    \end{tabular}
    \label{tab_model_compare}
\end{table}

\begin{tcolorbox}
\textbf{Finding 2-1.} 
The LLMs demonstrate a remarkable ability to surpass state-of-the-art models in both cross-domain and in-domain scenarios, achieving an improvement of 26\% and 16\% in terms of Exact Accuracy, respectively. This underscores the LLMs' exceptional generalization capabilities. Furthermore, with the provision of more few-shot samples, the performance of inference-only models consistently improves, eventually surpassing that of fine-tuned models.
\end{tcolorbox}

\subsubsection{RQ2-2: How does the number of in-context demonstrations affect the performance of LLMs?}
\label{RQ2-2}

In this RQ, we explore the impact of demonstration selection methods on in-context learning for \nltovis task, by selecting different databases and varying quantities of examples in demonstration.
To investigate the effect of the number of databases and examples per database in the demonstration, we conduct experiments encompassing various combinations. 
Specifically, our demonstration examples for in-context learning are drawn from $A$ distinct databases to simulate the cross-domain scenario. 
From each of these databases, we extract $B$ paired instances consisting of a natural-language query and its corresponding VQL query.
Collectively, this amounts to \( C = A \times B \) examples.
In our configuration, we permit both variables $A$ and $B$ to attain a maximum value of 4, ensuring that their cumulative total does not surpass the length limit specified by the prompt.
We conduct this experiment using the \textit{Table2SQL} prompt in a cross-domain scenario across all samples within the test dataset.

\begin{figure}[!t]
\centering
\includegraphics[width=0.7\textwidth]{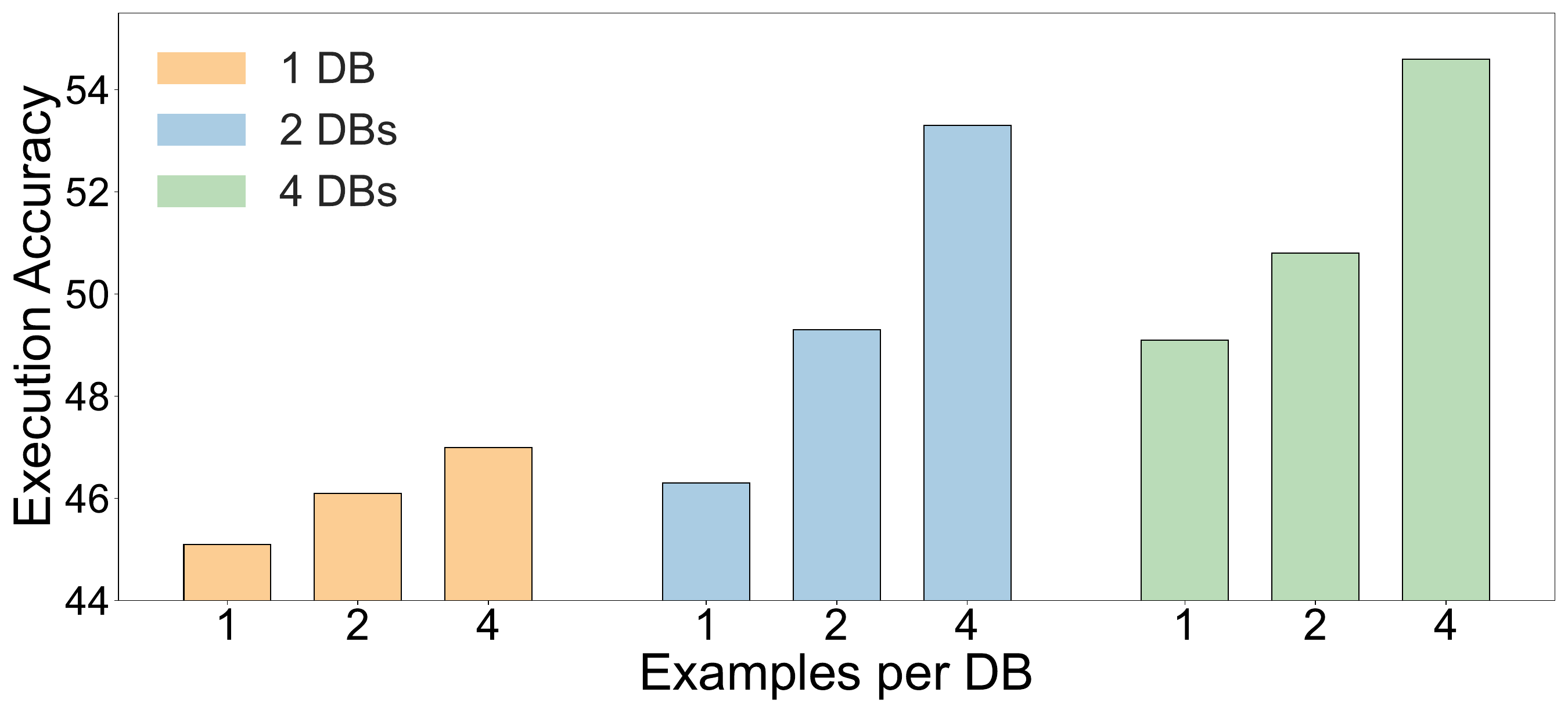}
\caption{
The average Execution Accuracy across test dataset using \textit{Table2SQL} by \texttt{text-davinci-003} with respect to different numbers of databases (DBs) and examples per database (Exp/DB) in the demonstration for in-context learning.
\label{fig_database_example}
}
\label{fig_RQ2_3}
\end{figure}

In Figure~\ref{fig_RQ2_3}, it is evident that there is a notable improvement (45\%-47\%) in performance when all examples are sourced from the same database. This improvement is observed as the number of examples increases from 1 to 4.
When the total number of examples is fixed (e.g., at 4), sourcing them from entirely different databases reveals superior performance (49\%) compared to sourcing them from the same database (47\%). It indicates that, in in-context learning, selecting a greater number of examples from diverse databases is beneficial.

\begin{tcolorbox}
\textbf{Finding 2-2.} 
Multiple cross-domain demonstrations are generally more beneficial than examples derived from the same database domain.
\end{tcolorbox}

\subsubsection{User Study}
\label{user_study}
We conduct a user study on LLMs to evaluate whether LLMs work well in the real world with users from different backgrounds. First, we design two tasks as follows: (1) Given tables and a target visualization with description, users express a natural-language query for creating such a visualization. (2) If the query can not generate the target visualization correctly, the users could choose to revise it three times. 
Next, we invite 3 graduate students as experts and 3 undergraduate students as non-experts, all majoring in computer science, to participate in the user study.
Each expert possesses over six years of proficiency in software development, demonstrating advanced skills in data analysis and visualization. On the other hand, non-experts, with approximately two years of programming experience, are capable of executing basic visualization operations in Excel.
Then, we select 5 databases at random to ensure a diverse range of data for our study. 
From each database, we pick 3 visualizations for each of the 4 difficulty levels (i.e., easy, medium, hard, and extra hard), resulting in a total of 60 visualizations.
We design a command-line interface that enables users to engage in data visualization over the selected databases using LLMs by crafting natural-language queries.
Subsequently, these crafted natural-language queries, along with the serialized table and a set of 20 demonstration examples, are input into the LLM (i.e., \texttt{text-davinci-003}) through in-context prompting.
The generated VQL queries will be transformed into visualizations for users, allowing them to iteratively revise the natural-language queries.

\begin{figure}[!t]
	\centering
        \includegraphics[width=0.6\textwidth]{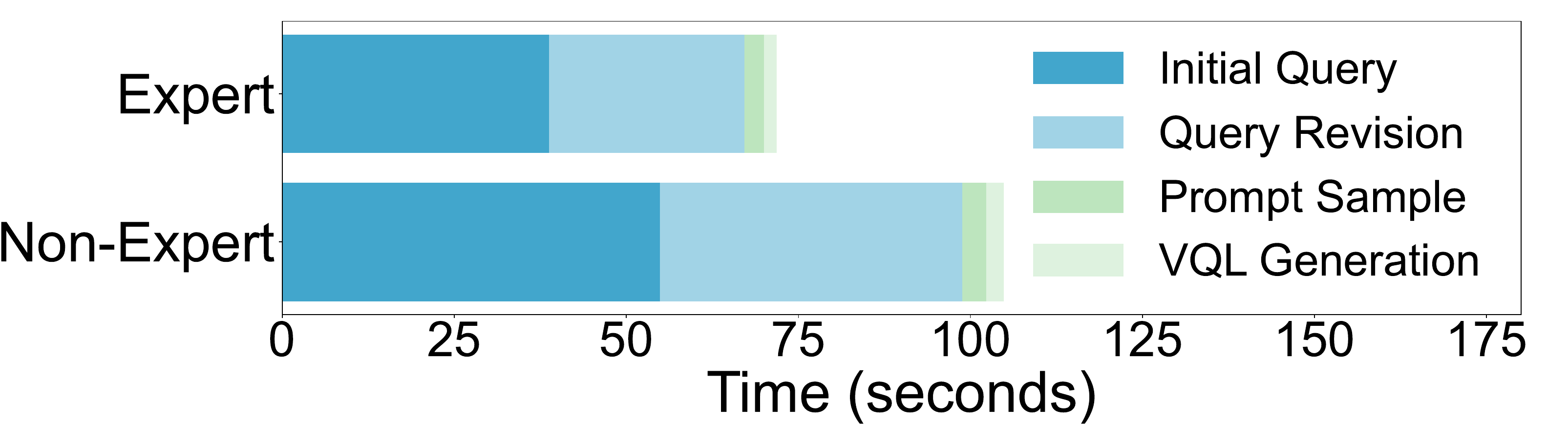}
\caption{
Quantitative analysis of the composition of user time on average.
}
\label{fig_time}
\end{figure}

\begin{figure}[!t]
\centering
\includegraphics[width=0.6\textwidth]{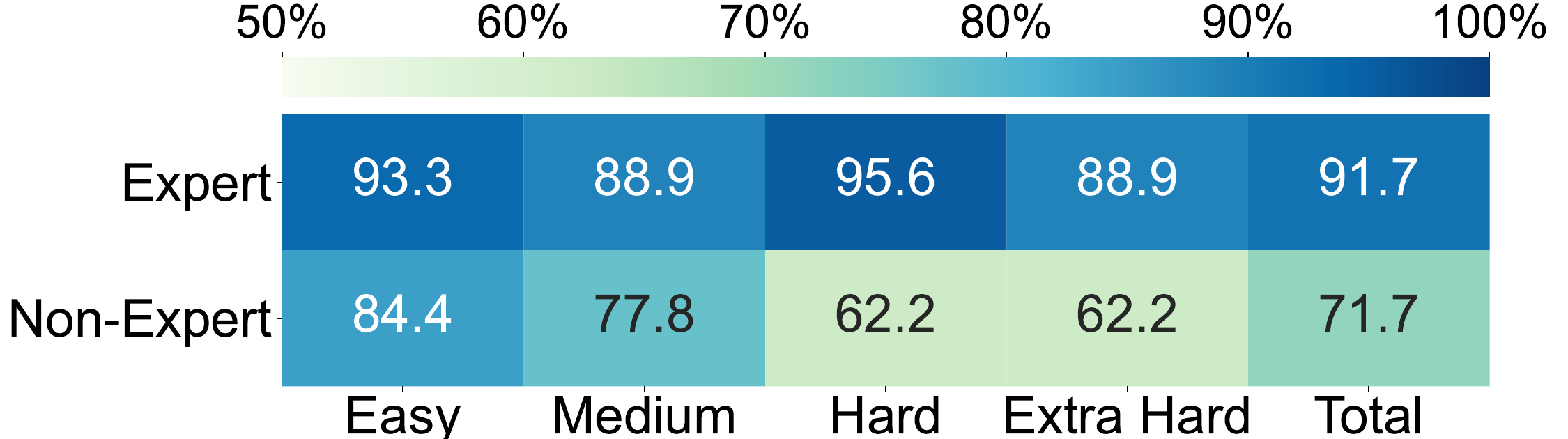}
\caption{
The average success rates of experts and non-experts queried for 4 difficult levels of visualization.
}

\label{fig_surpass}
\end{figure}

Figure~\ref{fig_surpass} shows the success rates of users querying for visualizations across four levels of difficulty. We observe that the experts are excellent in formulating queries for complex visualizations, successfully obtaining 95.6\% hard charts. 
While the non-experts are proficient in queries for 84.4\% easy charts.
Figure~\ref{fig_time} reports that non-experts take approximately 16 seconds more for the initial composition of queries and 15 seconds more for revision than experts. 
The system consistently maintains an average response time of 3 seconds for generating prompt examples and 2 seconds for VQL generation, applicable to both user groups.
Finally, we collect feedback on the ability of LLMs for the \nltovis task. All users acknowledge the LLM's proficient understanding of natural-language query for \nltovis task, and appreciate its substantial facilitation of the visualization process.

\subsection{RQ3: Iterative Updating}
\label{RQ3_result}

In this RQ, we commence by analyzing the conditions under which LLMs encounter failures in \nltovisnospace. Subsequently, we propose a strategy to mitigate these failures by iteratively updating the results through in-context learning with chain-of-thought.

\subsubsection{RQ3-1: When do the LLMs fail in \nltovisnospace?}
\label{RQ3-1}


To gain a deeper insight into the instances where LLMs encounter challenges, we conduct a comprehensive analysis of erroneous outputs. Specifically, we scrutinize all outcomes generated by the \texttt{text-davinci-003} model using the \textit{Table2SQL} prompt with a 20-shot approach in a cross-domain scenario. We classify the cases of failure into distinct categories, depending on the component (e.g., \texttt{wrong tables} and \texttt{wrong columns}) of the VQL it struggles to predict accurately.
Based on the AST of visualization query~\cite{luo2021nvbench}, a visualization consists of two components: the visual part (visualization types, axis) and the data part (transformation from a database)~\cite{luo2021synthesizing}. 
Specifically, the evaluation of visualization types involves measuring type tokens such as bar, scatter, line, and pie. With regard to the axis component, the assessment focuses on the ``\texttt{SELECT}'' component of the visualization query.
Data part includes ``\texttt{BIN}'', ``\texttt{GROUP}'', ``\texttt{JOIN}'', ``\texttt{COND (ORDER, WHERE, and AND/OR)}'', and nested components.

\begin{figure}[!t]
\centering
\includegraphics[width=0.65\textwidth]{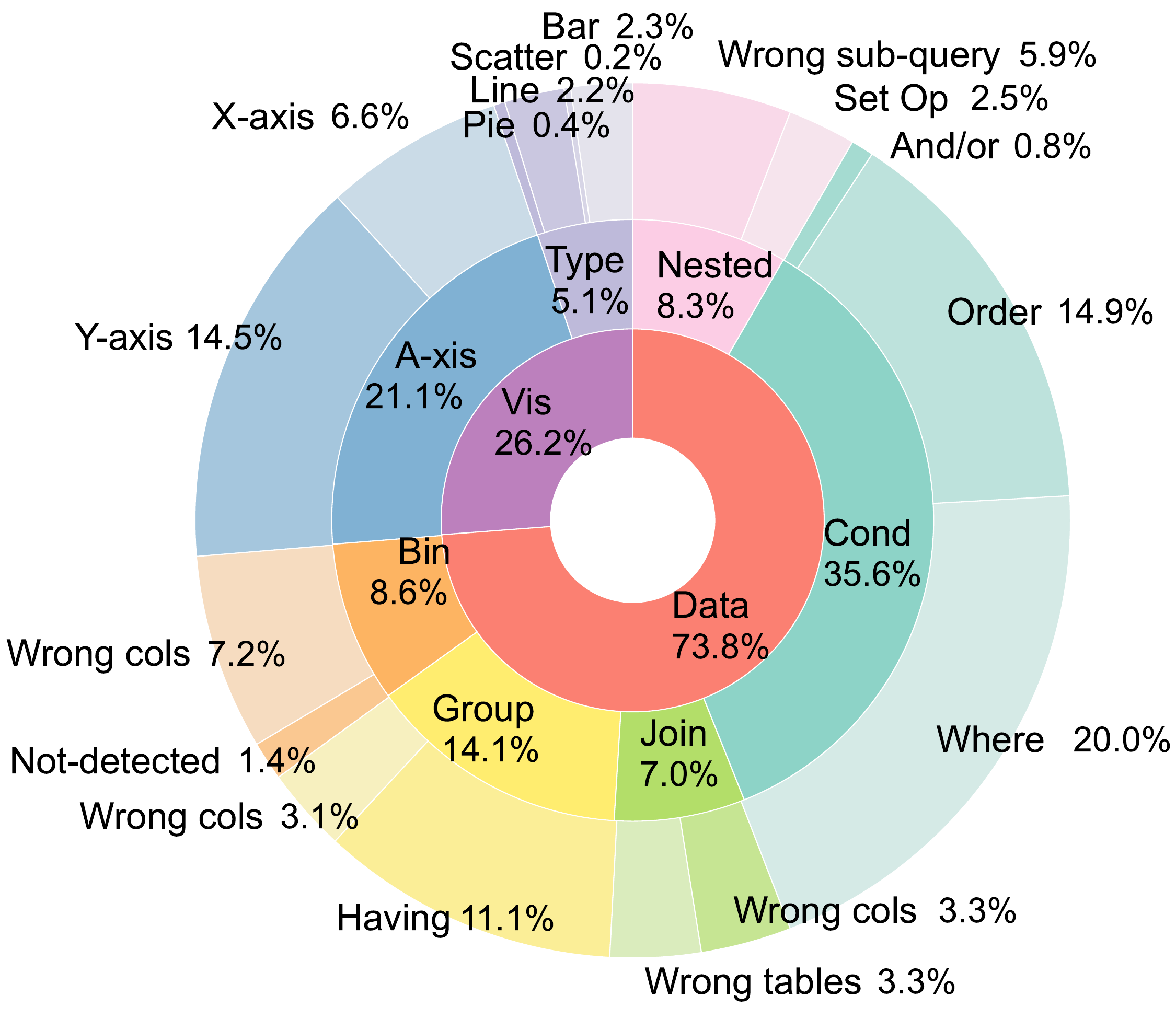}
\caption{Statistics of failures by \texttt{text-davinci-003} in 20-shot using \textit{Table2SQL}.
}
\label{fig_error}
\end{figure}

Figure \ref{fig_error} presents a breakdown of these failure statistics. It is evident from this figure that data-related errors constitute the majority at 73.8\%, which is approximately three times more prevalent than visual-related errors at 26.2\%. This observation underscores the challenge of precisely identifying and visualizing the relevant data.
In terms of the data-related errors, the highest proportion of errors is associated with the ``cond'' (condition) attribute (35.6\%). This indicates a need to enhance the ability of LLMs to effectively handle queries related to filtering operations. 
This error analysis motivates us to iteratively update the results in a conversational manner, mirroring the functioning of LLMs.

\begin{tcolorbox}
\textbf{Finding 3-1.} 
In summary, the analysis reveals that errors in the data part of the visualization query are more frequent compared to those in the visual part. Respectively, errors are mainly related to data filtering and the $y$-axis of visualization.
\end{tcolorbox}

\subsubsection{RQ3-2: Can we iteratively update the results via optimization strategies?\nopunct}
\label{RQ3-2}
\label{RQ_optimization}

\begin{figure}[!t]
\centering
\includegraphics[width=1\textwidth]{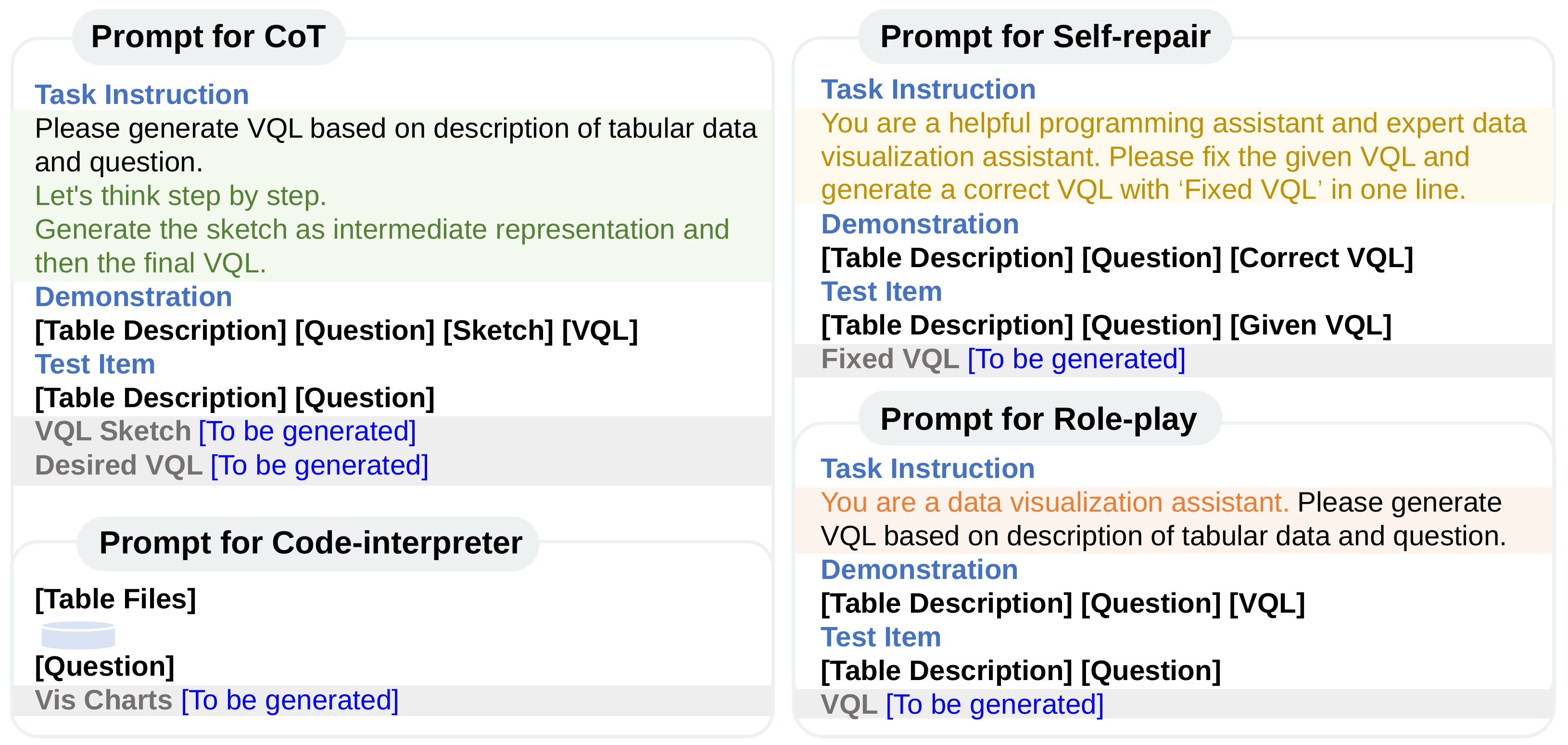}
\caption{
Optimization strategies employed for LLMs to iteratively update the results.
}
\label{figure_Optimization}
\end{figure}

Here, we employ the CoT strategy with manual construction~\cite{wei2022chain} to enhance the prompts for \nltovisnospace, infusing LLMs (i.e., \texttt{gpt-3.5-turbo} and \texttt{gpt-4}) with an intermediary cognitive process. 
Our approach utilizes a sketch as an intermediate expression, containing essential keywords from the visualization query. Additionally, we harness \texttt{gpt-3.5-turbo}'s innate self-instruct capability by introducing the phrase ``\textit{Let's think step by step.}'' This combined strategy is denoted as CoT.
Furthermore, we delve into role-playing, where \texttt{gpt-3.5-turbo} takes on the persona of a visualization expert. In consideration of \texttt{gpt-4}'s self-repair capabilities, we explore instructions for rectifying visualization queries. The optimization strategy is elucidated in Figure~\ref{figure_Optimization}. 
Additionally, we investigate GPT-4's newly introduced code-interpreter feature in ChatGPT Plus\footnote{\url{https://chat.openai.com/?model=gpt-4-code-interpreter}}, which enables GPT-4 to demonstrate programming proficiency within a conversational context, thus facilitating user learning and problem-solving.

We evaluate the total failure results of \texttt{text-davinci-003} with \textit{Table2SQL} in 20 shots to explore the optimization strategy.
For \texttt{gpt-3.5-turbo}, we employ the following strategies.
(1) CoT guided by the prompt ``\textit{Let's think step by step. Generate the sketch as an intermediate representation and then the final VQL}''. 
(2) Role-playing, where the prompt ``\textit{You are a data visualization assistant}'' sets the model's persona.
For GPT-4 (\texttt{gpt-4}), we explore the (3) self-repair by adopting the following prompt ``\textit{You are a helpful programming assistant and expert data visualization assistant. Please fix the given VQL and generate a correct VQL}''. 
We also explore the (4) code-interpreter by uploading database files and entering natural-language queries on the ChatGPT Plus website. By pairing every same VQL with only the first natural-language query as a test example, we filter 176 different charts from the LLM's failed dataset. 
The generated visualizations are manually checked for accuracy.

\begin{figure*}[!t]
    \begin{subfigure}[b][][c]{.48\textwidth}
	\centering
        \includegraphics[width=\linewidth]{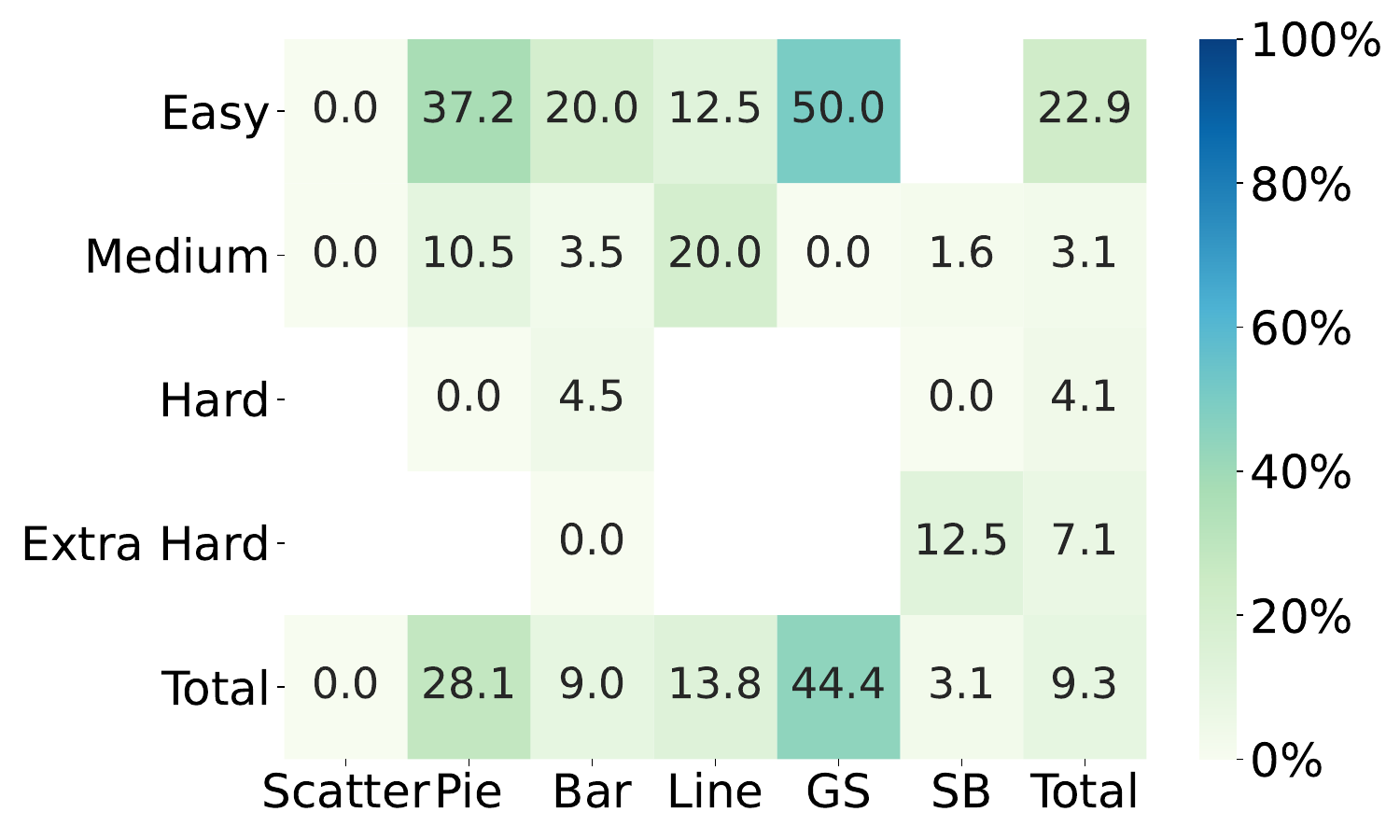}
	\caption{CoT}
	\label{fig_error_sketch}
    \end{subfigure}
    \begin{subfigure}[b][][c]{.48\textwidth}
	\centering
        \includegraphics[width=\linewidth]{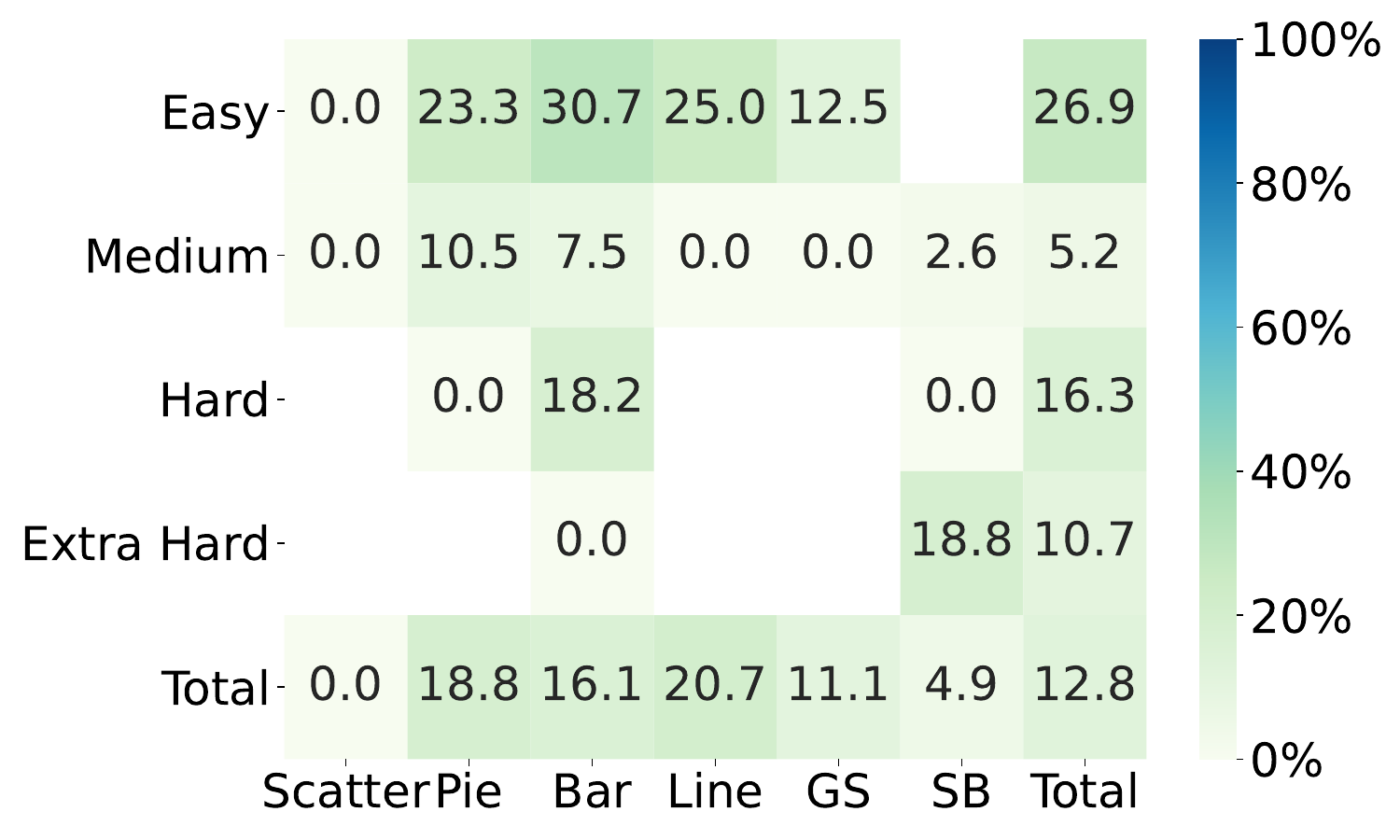}
	\caption{Role-playing}
	\label{fig_error_roleplay}
    \end{subfigure}
    \begin{subfigure}[b][][c]{.48\textwidth}
	\centering
        \includegraphics[width=\linewidth]{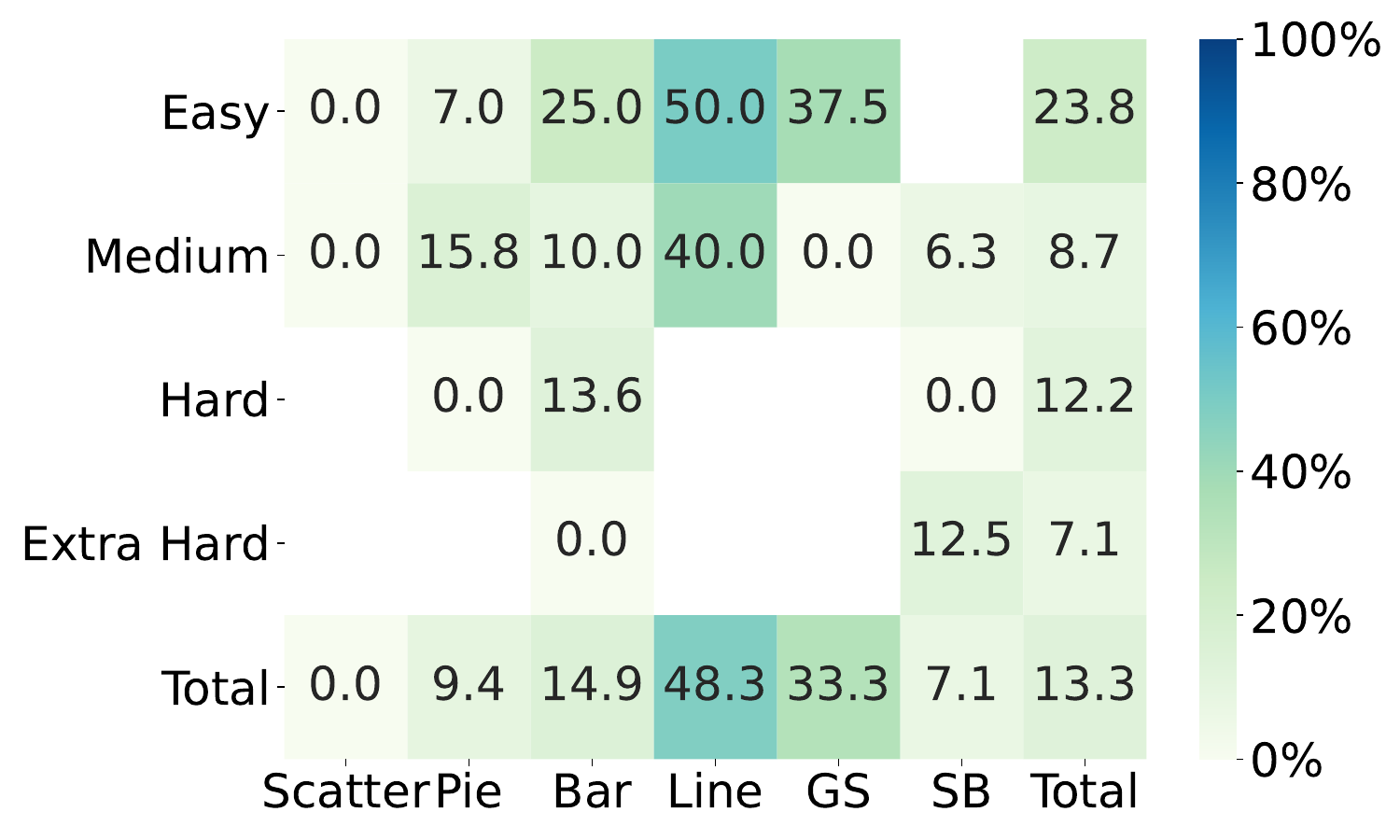}
	\caption{Self-repair}
	\label{fig_error_repair}
    \end{subfigure}
    \begin{subfigure}[b][][c]{.48\textwidth}
	\centering
        \includegraphics[width=\linewidth]{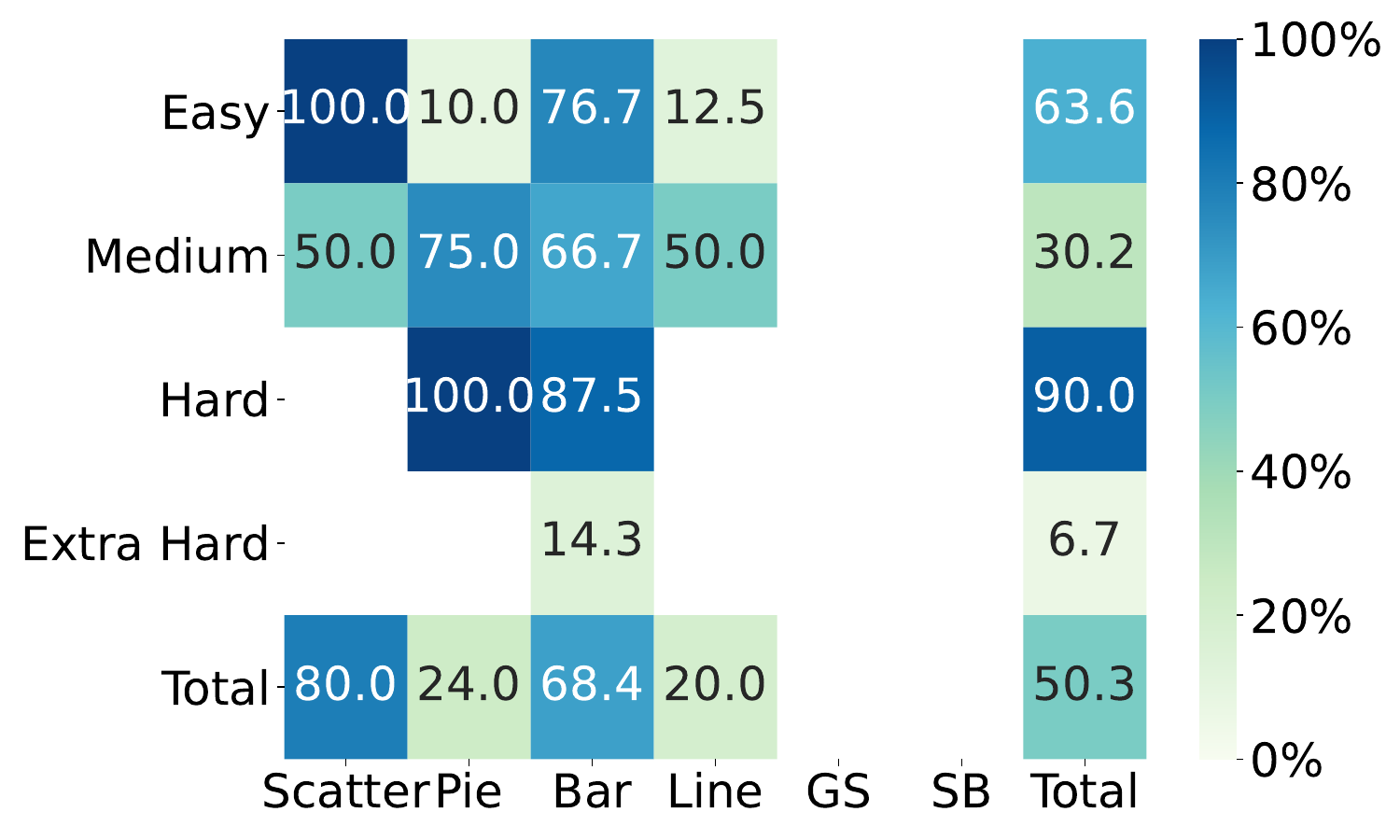}
	\caption{Code-interpreter}
	\label{fig_error_code_interpreter}
    \end{subfigure}
\caption{The Execution Accuracy of optimization strategy in the failed results of \texttt{text-davinci-003} in one-shot with \textit{Table2SQL}. GS and SB refer to grouping scatter and stacked bar chart types, respectively.}
\label{figure_Optimization_results}
\end{figure*}

Figure~\ref{figure_Optimization_results} depicts the comparative performance of diverse optimization strategies employed by both \texttt{gpt-3.5-turbo} and \texttt{gpt-4}.
Our findings reveal significant improvements in Execution Accuracy: the CoT strategy enhances accuracy by 9.3\%, while the role-playing strategy shows an impressive 12.8\% boost. Notably, \texttt{gpt-4}'s self-repair strategy leads to a remarkable 13.3\% accuracy improvement. 
When subjected to the code-interpreter on ChatGPT Plus within the extract dataset, the Execution Accuracy of visualization improves to 50.3\%.

\begin{tcolorbox}
\textbf{Finding 3-2.} 
The Self-repair strategy outperforms the CoT and role-playing strategies in updating the results within \nltovisnospace. Furthermore, the code-interpreter in \texttt{gpt-4} excels, indicating a promising avenue for future research.
\end{tcolorbox}

\section{Discussion}
\label{discussion}
In this section, we discuss our findings during the experiments and hope researchers can address some of the issues in future work.
\subsection{Findings and Implications}

In this study, we have uncovered several significant findings that offer valuable insights.

\smallskip
\noindent\textbf{Finding 1}: Investigating the strategies of inputting tables into LLMs, we find that representing tables in a programming language format is most effective, showcasing remarkable performance in addressing \nltovis tasks when working with structured tabular data. Future work should delve deeper into investigating the intricacies of table representation within programming languages. 

\smallskip
\noindent\textbf{Finding 2}: In analyzing the impact of three table components (i.e., schema, relationship, and content) in prompts, we identify that table schema is the most crucial component in both cross-domain and in-domain settings. Table content is found to be inconsequential in \nltovis tasks, emphasizing the importance of the format and relationships in future research, especially when feeding extra large databases into LLMs.

\smallskip
\noindent\textbf{Finding 3}: LLMs exhibit superior capabilities for automating data visualization based on natural-language descriptions, outperforming state-of-the-art models in both cross-domain and in-domain scenarios. Furthermore, as more few-shot samples are provided, inference-only models excel, even surpassing fine-tuned models. It is suggested to apply our framework to open-source LLMs for more general \nltovis task, which involves implicit and multi-type table structures.

\smallskip
\noindent\textbf{Finding 4}: Investigating the selection of demonstrations in in-context learning, we determine that multiple out-of-domain demonstrations generally offer greater benefits than examples from the same database domain, which could guide the design of demonstration selection in the future.

\smallskip
\noindent\textbf{Finding 5}: The analysis of failed generation reveals that errors are more frequent in the data part of the visualization query, with particular issues related to data filtering and the $y$-axis of visualization. There is room for improvement in generating condition and group attributes, which inspires further exploration of iterative updating based on multi-turn dialogs of LLMs.

\smallskip
\noindent\textbf{Finding 6}: Optimization strategies such as the self-repair of \texttt{gpt-4} and ChatGPT Plus's code-interpreter demonstrate superior performance, although challenges persist in handling complex natural-language queries and join cases. Future work should focus on intuitive interfaces to enhance the usability and reliability of \nltovis tasks in real-world applications.

\subsection{Limitations}
\label{threats}
There are some potential limitations on the validity of our experimental results and conclusions.

\noindentparagraph{\textbf{\textup{Limited tasks and dataset.}}}

In this work, we explore the \nltovis task with regular table structures and descriptive table context. However, in practice, tables could be in irregular structures, such as those featuring merged rows and columns in Excel, or containing mixed data content. On the other hand, table schema and column headers may not be descriptive. Future studies can extend the dataset to irregular tables for visualization, to enrich \nltovis benchmarks in more applicable scenarios.
Moreover, all experiments in this paper are assessed using a synthesized dataset derived from the \textsc{NL2SQL} dataset. 
We anticipate extending our study to real-world datasets in future work.

\noindentparagraph{\textbf{\textup{Limited visualization specification.}}}

In this paper, we choose VQL, an intermediate language that has been widely embraced in existing literature. 
While some may contend that visualizations can be directly generated through the generation of high-level specifications, such as Vega-Lite (in JSON format), we argue that this approach may encounter difficulties in accurately capturing specific details, such as chart type and rendering color.
As part of our future work, we plan to explore the direct generation of diverse Vega-Lite specifications. 

\noindentparagraph{\textbf{\textup{Support of conversational \nltovisnospace.}}}
Data analysts typically perform data visualization in a conversational manner. Take conversational visual analysis for example, a conversational natural-language inquiry can be made up of a number of separate but related natural-language inquiries. It is an intriguing and promising avenue to extend the \nltovis benchmark to conversational visual data analytics.

\noindentparagraph{\textbf{\textup{Manual prompt design.}}}
Like other prompting strategies, the prompt design and optimizations are based on human comprehension and observations. The designer's knowledge may affect the effectiveness of the used prompts. Nonetheless, the purpose of our study is to investigate the viability of prompt design and related influential factors. Our experimental results demonstrate that the designed prompts are effective. In the future, we will investigate automated prompt construction techniques~\cite{liu2023pre}.
 
\section{Related Work}
In this section, we review the related literature about \nltovis task, LLMs for code generation, and LLMs for data engineering.
\label{related_work}

\subsection{\nltovisnospace}

\nltovis is crucial in data analysis, aiming to generate visualization representations from brief succinct natural language explanations. While many approaches harnessing rule-based NLP techniques have been proposed~\cite{gao2015datatone,setlur2016eviza,hoque2017applying} to address this challenge, their effectiveness is limited by the inflexibility of predefined rules in handling open-form natural-language inputs.
Owing to the availability of large-scale corpora of natural-language descriptions paired with corresponding visualizations, such as nvBench~\cite{luo2021nvbench}, deep-learning-based techniques~\cite{luo2018deepeye,luo2021synthesizing,luo2021nvbench,luo2021natural} have been employed to train a sequence-to-sequence model in an end-to-end manner for the \nltovis task. For example, \textsc{Seq2Vis}~\cite{luo2021synthesizing} utilizes an LSTM to encode natural-language queries into hidden states, which are subsequently decoded into visualizations. Moreover, ncNet~\cite{luo2021natural} considers another encoder-decoder architecture, the Transformer~\cite{vaswani2017attention}, to translate natural-language descriptions into visualizations.
Leveraging the powerful capabilities of ChatGPT~\cite{openai2022chatgpt}, Chat2Vis~\cite{maddigan2023chat2vis} invokes the API interfaces of \texttt{code-davinci-002} to enable users to create data visualizations using natural-language queries in Python plots. 
More recently, RGVisNet~\cite{song2022rgvisnet} has introduced a retrieval-based model, designed to retrieve the most pertinent visualization representation from an extensive visualization codebase for a given natural-language query. 

\subsection{LLMs for Code Generation}
Code generation targets the automatic generation of program code from natural-language descriptions,
which can assist developers in improving programming productivity and efficiency. 
In recent years, LLMs have exhibited remarkable capabilities in generating code~\cite{bommasani2021opportunities}, including Python programs, execution commands for Excel, and SQL queries for databases.
These models are generally built upon the Transformer architecture and pre-trained on large-scale corpora using self-supervised objectives such as masked language modeling and next-sentence prediction~\cite{lee2018pre}. 
Examples of these models include CodeGen~\cite{nijkamp2022codegen}, CodeT5+~\cite{wang2021codet5}, ChatGPT~\cite{openai2022chatgpt}, StarCoder~\cite{li2023starcoder}, and Code Llama~\cite{roziere2023code}.
To fully harness the zero-shot potential of LLMs, a range of techniques have emerged, including prompt tuning, in-context learning, chain-of-thought, and instruction tuning.
In particular, in-context learning stands out as a method to fortify LLMs by providing contextual information or illustrative examples, as explored in~\cite{li2023large} for code generation. Similarly, chain-of-thought is devised to ensure the logical coherence of LLM outputs, thereby enhancing the performance of code generation~\cite{li2023enabling}. 
Moreover, instruction tuning has been conceived to enhance the generalization prowess of LLMs across various tasks, exemplified by the creation of WizardCoder~\cite{luo2023wizardcoder}, which augments the capabilities of StarCoder through the innovative \textit{Evol-Instruct} approach to generate sophisticated code instructions.

\subsection{LLMs for Data Engineering}
In the field of data engineering and analysis, the integration of LLMs with data-centric tasks has opened avenues for transformative approaches to data interaction, processing, and visualization. 
Recently, numerous studies have been developed to weave natural language into tabular data analysis~\cite{hu2023chatdb,li2023graphix,zhang2023data,li2023sheetcopilot,zha2023tablegpt}. For instance, \textsc{NL2SQL}~\cite{hu2023chatdb,li2023graphix} adeptly translates natural language into SQL commands to manipulate relational databases. Additionally, ChatExcel~\cite{ChatExcel} and \textsc{NL2Formula}~\cite{zhao2024nl2formula} have leveraged LLMs to generate Excel execution commands, thereby streamlining user interactions. SheetCopilot~\cite{li2023sheetcopilot} has explored translating languages to VBA (\textit{Visual Basic for Applications} - an embedded scripting language in Microsoft Excel), benefiting from a rich array of spreadsheet software functionalities. Data-Copilot~\cite{zhang2023data}, an LLM-based system, facilitates the automated management, invocation, and processing of a substantial volume of data from various sources, crafting sophisticated interface tools autonomously. 
Designed for table analysis, TableGPT~\cite{zha2023tablegpt}, 
blends tables, natural language, and commands to manipulate data, visualize information, generate analysis reports, answer questions, and make predictions.

\section{Conclusion and Future Work}
\label{conclusion}
In this paper, we have investigated whether it is feasible to utilize LLMs for the \nltovis task.
Specifically, we compare LLMs including fine-tuned models (e.g., T5-Small, T5-Base) and inference-only models (e.g., \texttt{text-davinci-002}, \texttt{text-davinci-003}, \texttt{gpt-3.5 -turbo}, and \texttt{gpt-4}) against the state-of-the-art models. 
To start with, we investigate different approaches to transforming the structured tabular data into sequential prompts, so as to feed them into LLMs. 
Furthermore, we evaluate the LLMs on the \nltovis benchmark under in-domain and cross-domain settings, against several traditional neural models for \nltovisnospace.
At last, we analyze when the LLMs fail in \nltovisnospace, and propose to iteratively update the results using strategies such as chain-of-thought, role-playing, and code-interpreter.

In our future work, we plan to explore table representations by encoding tables with more well-designed programming languages. Moreover, we will extend the benchmarks to implicit and multi-type table structures for visualization in more applicable scenarios, to push forward the field of \nltovisnospace.
While our current focus is on few-shot \nltovis in GPT-3.5, our framework can also be applied to other LLMs, such as LLaMA~\cite{touvron2023llama}. Consequently, the application of our prompting methods to multiple semantic parsing tasks represents an intriguing avenue for future exploration. 

\begin{acks}
This work is supported by the Major Program (JD) of Hubei Province (Grant No. 2023BAA024).
We would like to thank all the anonymous reviewers for their insightful comments.
\end{acks}

\bibliographystyle{ACM-Reference-Format}
\bibliography{ref}

\end{document}